\documentclass[iop]{emulateapj}
\usepackage{apjfonts}
\pdfoutput=1

\usepackage[bookmarks=true, colorlinks=true, citecolor=blue,backref=true]{hyperref}

\newcommand{\msperyr}{{M_{\odot}{\rm yr}^{-1}}}

\newcommand{\ha}{$\rm{H}\alpha$}
\newcommand{\hb}{$\rm{H}\beta$}
\newcommand{\mic}{\mu{\rm m}}

\newcommand{\galex}{{\it GALEX}}
\newcommand{\wise}{{\it WISE}}

\slugcomment{Draft}
\journalinfo{Submitted to Astrophysical Journal Supplement Series}

\shorttitle{GALEX--SDSS--WISE LEGACY CATALOG}
\shortauthors{SALIM ET AL.}

\begin{document}

\title{GALEX--SDSS--WISE Legacy Catalog (GSWLC):\\ Star Formation
  Rates, Stellar Masses and Dust Attenuations of 700,000 Low-redshift
  Galaxies}

\author{Samir Salim\altaffilmark{1}, Janice C.\ Lee\altaffilmark{2,3},
 Steven Janowiecki\altaffilmark{4}, Elisabete da Cunha\altaffilmark{5}, Mark
 Dickinson\altaffilmark{6}, M\'ed\'eric Boquien\altaffilmark{7}, Denis
 Burgarella\altaffilmark{8}, John J.\  Salzer\altaffilmark{1}, St\'ephane Charlot\altaffilmark{9}}
\altaffiltext{1}{Department of Astronomy, Indiana University,
   Bloomington, IN 47404, USA} 
\altaffiltext{2}{Space Telescope Science Institute, Baltimore, MD 21218, USA} 
\altaffiltext{3}{Visiting Astronomer, Spitzer Science Center, Caltech,
   Pasadena, CA 91125, USA}
\altaffiltext{4}{International Centre for Radio Astronomy Research
  (ICRAR), The University of Western Australia, Perth, WA 6009,
  Australia}
\altaffiltext{5}{Research School of Astronomy and Astrophysics,
  Australian National University, Canberra, ACT 2611, Australia}
\altaffiltext{6}{National Optical Astronomy Observatory, Tucson, AZ
   85719, USA}
\altaffiltext{7}{Universidad de Antofagasta, Unidad de Astronom\'ia,
  Antofagasta 1270300, Chile}
\altaffiltext{8}{Laboratoire d'Astrophysique de Marseille,
  Aix-Marseille Universite,  CNRS, F-13013 Marseille, France}
\altaffiltext{9}{Institut d'Astrophysique de Paris, CNRS, F-75014 Paris, France}
\email{salims@indiana.edu}

\begin{abstract}
  In this paper, we present GALEX--SDSS--WISE Legacy Catalog (GSWLC),
  a catalog of physical properties (stellar masses, dust attenuations
  and star formation rates (SFRs)) for $\sim\!700,000$ galaxies with
  SDSS redshifts below 0.3. GSWLC contains galaxies within the \galex\
  footprint, regardless of a UV detection, covering 90\% of SDSS. The
  physical properties were obtained from UV/optical SED fitting
  following Bayesian methodology of \citet{s07}, with improvements
  such as blending corrections for low-resolution UV photometry,
  flexible dust attenuation laws, and emission line corrections.
  GSWLC includes mid-IR SFRs derived from IR templates based upon 22
  $\mic$ \wise\ observations. These estimates are independent of
  UV/optical SED fitting, in order to separate possible
  systematics. The paper argues that the comparison of specific SFRs
  (SSFRs) is more informative and physically motivated than the
  comparison of SFRs.  SSFRs resulting from the UV/optical SED fitting
  are compared to the mid-IR SSFRs, and to SSFRs from three published
  catalogs. For ``main sequence'' galaxies with no AGN contribution
  all SSFRs are in very good agreement (within 0.1 dex on average).
  In particular, the widely-used aperture-corrected SFRs from MPA/JHU
  catalog show no systematic offsets, in contrast to some
  integral-field spectroscopy results.  For galaxies below the main
  sequence (log SSFR$<-11$), mid-IR (S)SFRs based on fixed
  luminosity--SFR conversion are severely biased (up to 2 dex) because
  the dust is primarily heated by old stars. Furthermore, mid-IR
  (S)SFRs are overestimated by up to 0.6 dex for galaxies with AGN,
  presumably due to non-stellar dust heating. UV/optical (S)SFRs are
  thus preferred to IR-based (S)SFRs for quenched galaxies and those
  which host AGN.
\end{abstract}


\section{Introduction}

The stellar mass and the current star formation rate (SFR) are two of
the most fundamental physical properties of a galaxy.  The SFR
normalized by stellar mass, i.e., the specific star formation rate
(SSFR, \citealt{tully82,bothun82}), in addition serves as a rough
indicator of a galaxy's star formation (SF) history. To infer
properties such as stellar mass and SFR from observed quantities,
stellar populations and the effects of dust attenuation are modeled
and compared to galaxy spectra (e.g., \citealt{cidfernandes05}) and/or
the integrated photometry.

Pioneering work to derive physical properties of galaxies from their
integrated light (e.g.,
\citealt{faber72,searle73,tinsley76,larsontinsley78}) paved the way
for the development of more advanced stellar population synthesis
(SPS) models in the 1980s (e.g.,
\citealt{bruzual83,renzini86,guiderdoni87}). These conventional models
were replaced by isochrone synthesis in the following decade (e.g.,
\citealt{cb91,bc93}), culminating in modern, high-resolution models
(e.g., \citep{bc03,m05,conroy10b}).

SPS models are the basis for widely used ``simple'' relations for
inferring the stellar mass from broad-band optical colors (e.g.,
\citealt{bell01,courteau14}). These relations are by necessity
averaged over galaxy populations and use simplified assumptions, so
they cannot take into account variations in SF history, dust
attenuation or stellar metallicity. Simple relations are also used for
inferring the SFR from UV or \ha\ luminosity
\citep{kennicutt98,calzetti13}. In addition to again using average
assumptions, these luminosities first need to be corrected for dust
attenuation.  For galaxies at $z\gtrsim 0$, the quantities that enter
these relations must be explicitly corrected to rest-frame
luminosities, which requires the spectral energy distribution (SED) of
a galaxy to be assumed.

A more flexible and comprehensive approach is to derive physical
parameters through SED fitting (see \citealt{walcher11,conroy13} for
overview). SPS models are used to produce a library (a grid, in case
of regular sampling intervals) of models in which the SF history and
metallicity take a range of values. Model SEDs are further subjected to
varying degrees of dust attenuation assuming some attenuation
law. Redshifted model fluxes (broad-band magnitudes) are compared to
observations (obviating a need for an explicit K-correction), and the
best-fitting model is sought and its parameters are adopted as the
parameters of the observed galaxy. Unlike with simple relations, in
SED fitting every observed flux plays some role in constraining all
derived parameters, taking full advantage of the full set of
observations.  While in SED fitting the relationships between the
observed and derived parameters are not explicit as with the simple
relations, it should be noted that both approaches are ultimately
rooted in SPS models.

For stellar SED fitting to produce useful constraints on the ongoing
SFR, the rest-frame UV photometry must be included.  SED fitting was
initially employed to derive SFRs from broad-band observations of
small samples of Lyman break galaxies at $2<z<3$, for which the UV was
redshifted to easily observed optical range
\citep{sawicki98,papovich01,shapley05}.  Subsequently, the UV surveys
carried out by {\it Galaxy Evolution Explorer} (\galex) satellite
\citep{martin05} placed low-redshift ($z<0.3$) galaxies within the
domain of SED fitting.  \citet{s05,s07} performed the SED fitting of
low-redshift galaxies by combining the UV photometry from \galex\ with
the optical photometry from Sloan Digital Sky Survey
(SDSS). Furthermore, they utilized a Bayesian approach to SED fitting,
following the methods pioneered by the Max Planck Institute for
Astrophysics / Johns Hopkins University (MPA/JHU) group
\citep{k03a,b04,tremonti04}.  Unlike the traditional best-fit (minimum
$\chi^2$) approach, Bayesian SED fitting determines the full
probability distribution function of any parameter (or combination of
parameters), yielding more robust parameter characterization, along
with its uncertainty (e.g., \citealt{taylor11}). Subsequent efforts
expanded Bayesian SED fitting to include thermal dust emission
\citep{dacunha08,noll09} and star formation histories derived from
cosmological simulations \citep{pacifici13}.

\citet{s07} (hereafter S07) performed a thorough investigation of SFRs
in the local universe and showed that the UV/optical SED fitting is
especially powerful for obtaining SFRs of galaxies with low
SSFRs. However, S07 used an early \galex\ data release covering only
10\% of SDSS area. Because of its preliminary nature, the catalog of
physical parameters from S07 was not released to the public.

Here we present GALEX--SDSS--WISE Legacy Catalog (GSWLC) of physical
parameters, which builds on S07 UV/optical SED fitting efforts, but
includes numerous methodological improvements, such as UV photometry
corrections, flexible dust attenuation curves, and emission line
corrections. Many of these improvements were made possible by the use
of the Code Investigating GALaxy Emission
(CIGALE).\footnote{\url{http://cigale.lam.fr}} (\citealt{noll09}, M.\
Boquien et al.\, in prep.), to calculate the models and perform the
SED fitting. GSWLC contains SDSS galaxies within the \galex\ footprint
(regardless of the UV detection), covering 90\% of SDSS area. In
addition to the stellar masses and the SFRs from the UV/optical SED
fitting, GSWLC also includes SFRs derived independently from {\it The
  Wide-field Infrared Survey Explorer} (\wise, \citealt{wise})
mid-infrared (IR) observations. Community access to the catalog is
provided online at \url{http://pages.iu.edu/~salims/gswlc} and from
Mikulski Archive for Space Telescopes (MAST).

After defining the scope and the sample of GSWLC (Section
\ref{sec:sample}), we describe the input data and the process of
deriving the parameters (Sections \ref{sec:data}, \ref{sec:sed} and
\ref{sec:wise}). The catalog is described in Section
\ref{sec:catalog}. In Section \ref{sec:comp} we examine the contents
of the catalog, while in Section \ref{sec:other} GSWLC is compared to
several previously published catalogs of physical properties of SDSS
galaxies.

\begin{figure}
\epsscale{0.9} \plotone{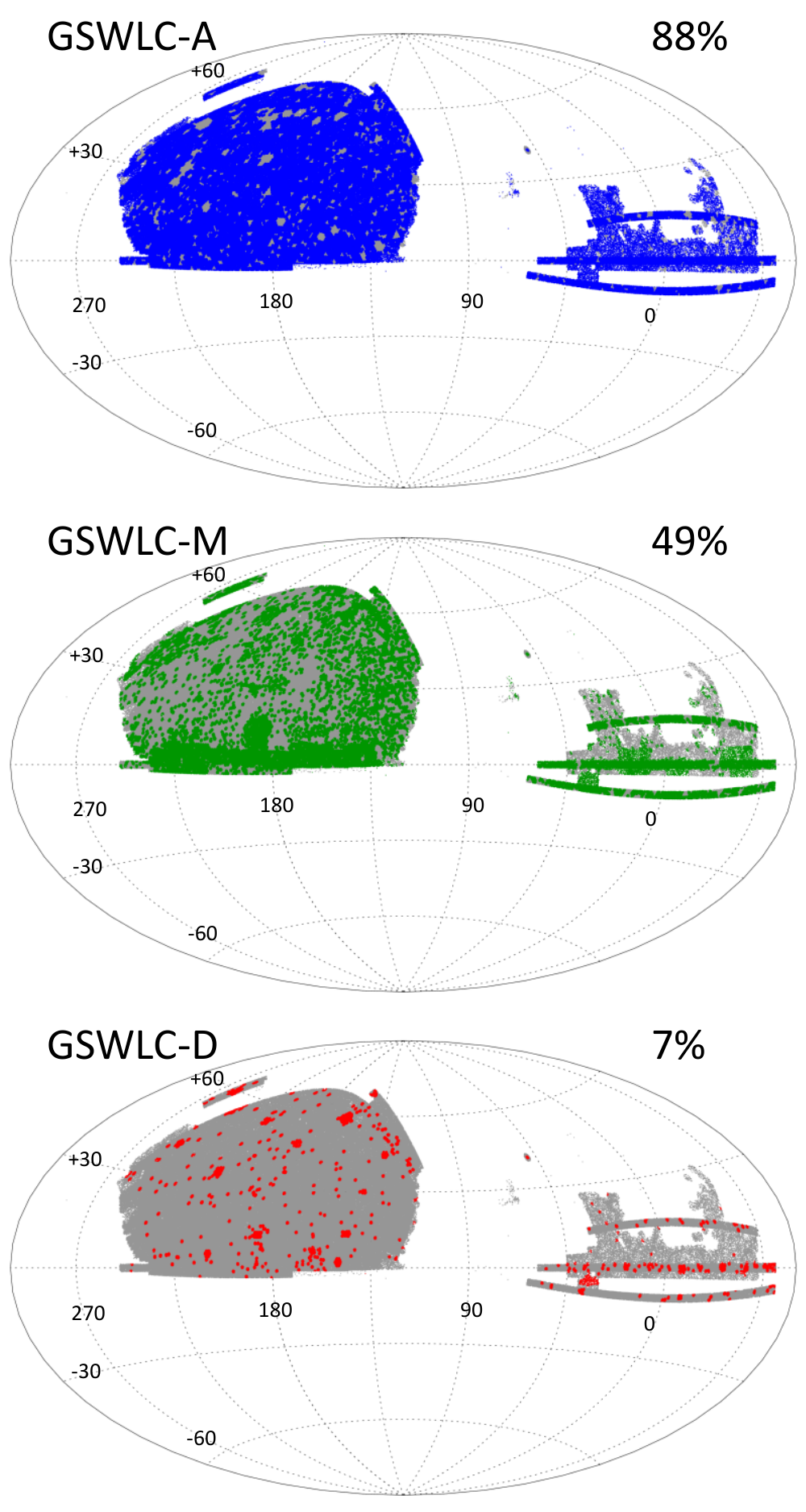}
\caption{Distribution on the sky of galaxies included in
  GALEX--SDSS--WISE Legacy Catalog (GSWLC). Each panel shows SDSS
  target galaxies as shaded areas, covering $\sim\! 9000$ sq.\
  deg. The SDSS Main Galaxy Sample (MGS) legacy survey area consists
  of the contiguous area on the left (Northern Galactic Cap) and three
  horizontal stripes on the right. Non-stripe areas on the right are
  primarily from SDSS BOSS survey of passive galaxies. BOSS galaxies
  are also interspersed in the contiguous area on the
  left. Superimposed on each panel in color are SDSS galaxies that
  fall within the coverage of one of \galex\ imaging surveys: shallow
  (all-sky) survey (AIS), medium-deep survey (MIS) and the deep survey
  (DIS), thus defining the coverage of GSWLC-A, M and D. Percentage of
  SDSS targets covered by each \galex\ survey is indicated in the
  upper right corner of each panel. \label{fig:map}}
\end{figure}

\section{Sample} \label{sec:sample}

%
\subsection{Scope and SDSS target sample} \label{ssec:scope}

GSWLC is built around SDSS Main Galaxy Sample (MGS, \citealt{sdsssp}),
the most extensive spectroscopic survey of low-redshift ($z<0.3$)
universe. More specifically, GSWLC includes all MGS-like SDSS galaxies
that fall within \galex\ footprint, covering $\sim\! 90\%$ of SDSS
area, or $\sim\! 8000$ sq.\ deg. Galaxies are kept in the sample
regardless of the detection in the UV, thus retaining the optical
selection. SDSS and \galex\ data were obtained through SDSS CasJobs
SQL server.

The majority of low-redshift galaxies in SDSS were targeted
spectroscopically as part of the magnitude-limited MGS, one of the
three original (``legacy'') spectroscopic surveys (others being the
luminous red galaxy, LRG, and quasar surveys). Legacy surveys were
completed in 2008 with the release of DR7 \citep{sdssdr7}, and cover a
large contiguous area in the Northern Galactic cap, and three separate
stripes in the Southern Galactic cap (Figure \ref{fig:map}). These
areas lie almost entirely in the Northern sky ($\delta>-10\degr$), and
mostly at high Galactic latitude ($|b|>30\degr$).

MGS selection algorithm targets non-stellar objects with well-measured
photometry brighter than $r_{\rm petro}=17.77$, with some additional
cuts based on surface brightness \citep{sdsssp}. The magnitude limit
of MGS yields a sample of galaxies peaking at $z\sim0.1$, with very
few galaxies above $z=0.3$.

Selection of SDSS targets is performed on DR10 \citep{sdssdr10}, based
on the {\tt SpecPhoto} table, which combines primary (non-duplicate)
photometric objects with primary spectroscopic
observations. Nevertheless, in 25 cases more than one primary spectrum
was found for the same photometric object, in which case we selected the
spectrum that was closest to the photometric position, leaving 730,288
unique objects (objets with unique {\tt ObjID}). These SDSS targets
constitute shaded areas in Figure \ref{fig:map}.

Our selection for inclusion in GSWLC consists of only two criteria:

\begin{equation}
\begin{array}{c}
r_{\rm petro}<18.0 \\
0.01<z<0.30
\end{array}
\end{equation}

The first criterion replicates MGS brightness cut, with some rounding
to allow for targeting photometry fluctuations. The second cut removes
stars and very nearby galaxies whose photometry is likely to be less
accurate because of their large angular size \citep{west10}, and whose
redshifts are poorer indications of the distance
\citep{cflows3}. GSWLC therefore includes the entire MGS, complemented
by MGS-like galaxies from other programs that fall within the
brightness (and, effectively, the redshift) cut of MGS: 5.2\% from
BOSS (typically ellipticals), and the remaining 2.1\% from other
surveys (QSO, SEGUE). For studies that focus on general population of
galaxies, it is strongly recommended to use only the MGS galaxies
({\tt flag\_mgs} = 1 in GSWLC).

\subsection{The final sample}\label{ssec:sample}

GSWLC includes all SDSS targets within areas covered by \galex,
regardless of a UV detection. \galex\ has observed the sky in 1.2
deg.\ wide circular fields (``tiles''), with a wide range of exposure
times \citep{morrissey07}. The greatest UV sky coverage is provided by
shallow ($\sim\! 100$ s) observations, most of which were taken as
part of the All-sky Imaging Survey (AIS). \galex\ could not safely
point in the direction of bright stars, so even the All-sky survey's
coverage contains holes. Observations of medium depth, corresponding
to one \galex\ orbit ($\sim\! 1500$ s), are the basis for Medium
Imaging Survey (MIS), which specifically targeted areas within SDSS
footprint. For select fields, deeper observations were obtained by
co-adding the observations from multiple orbits, to produce the Deep
Imaging Survey (DIS), with nominal exposure time of 30,000 s.

Since the accuracy of UV photometry, and consequently of the derived
physical properties, will depend on the depth of the observations, we
produce three separate catalogs, each approximately corresponding to
shallow (all-sky; A), medium (M), and deep (D) UV imaging surveys. The
catalogs are designated as GSWLC-A, GSWLC-M, GSWLC-D. Each catalog is
based only on \galex\ imaging of certain depth, even if deeper
exposures of an object exist:

\begin{equation}
\begin{array}{ll}
{\rm GSWLC-A}: & t_{\rm NUV}\leq 650 \\
{\rm GSWLC-M}: & 650<t_{\rm NUV}<4000 \\
{\rm GSWLC-D}: & t_{\rm NUV}\geq 4000 \\ 
\end{array}
\end{equation}

\noindent where the exposure times in NUV band are given in
seconds. FUV exposure times are typically identical to NUV times (the
two bands were observed simultaneously), except in cases when the FUV
image was missing due to camera malfunction. Note that in GSWLC-M we
also include individual visits used for DIS co-adds, since
the exposure times of individual visits fall in the range for GSWLC-M.

To define the final samples for inclusion in the catalogs, we take all
SDSS targets from Section \ref{ssec:scope} that fall within 0.6 deg of
\galex\ tile centers of the specific UV survey (A, M or D). This gives the
following final sample sizes:

\begin{equation}
\begin{array}{lll}
{\rm GSWLC-A}: & 640,659 & (88\%) \\
{\rm GSWLC-M}: & 361,328 & (49\%) \\
{\rm GSWLC-D}:  & 48,401 & (7\%)
\end{array}
\label{eq:samplesize}
\end{equation} 

\noindent where the percentage indicates the fraction of all SDSS DR10
targets. GSWLC-M contains 7$\times$ as many galaxies as the MIS sample
used in S07.  Since \galex\ surveys are nested, objects in the deeper
catalogs are mostly included in the shallower ones. The number of
unique galaxies encompassed by the three catalogs, 658,911, is
therefore only slightly larger than the number of objects in GSWLC-A,
and corresponds to 90\% of SDSS target sample.

\subsection{SDSS-\galex\ matching}\label{ssec:match1}

\galex\ data are taken from the final data release (GR6/7)
\footnote{\url{http://galex.stsci.edu/GR6/}}. Matching of SDSS to
\galex\ is in general a non-trivial task \citep{budavari09} because of
the changes in galaxy morphology with wavelength, and different
resolutions ($5\arcsec$ for \galex\ vs.\ $1\farcs3$ for
SDSS). Furthermore, the \galex\ data release contains detections of
same objects from multiple tiles. Thus, one needs to address the cases
when multiple UV candidates exist for an optical source, and when the
UV source may be a blend of two or more optical sources.

In our case, the problem of SDSS--\galex\ matching is significantly
alleviated by the fact that it involves relatively bright objects,
having low sky density. In particular, we find that there are
essentially no cases of genuine multiple UV candidates within
$5\arcsec$ search radius.  The choice of search radius is based on
\galex\ positions of isolated SDSS targets, for which we find
negligible bulk offset with respect to SDSS ($0\farcs2$ in either RA
or Decl.), and $1\sigma$ 1D positional uncertainty of $1\farcs3$. When
multiple UV candidates are nominally present, they are either due to
multiple observations (overlapping tiles), or from NUV and FUV
detections of the same object that were erroneously left as separate
sources in \galex\ merged-band catalog. When there are multiple
observations of the same object, we take the one from the tile with
the longest FUV exposure (if no candidate has an available FUV image,
the tile with the longest NUV exposure is selected).

While each SDSS object has an unambiguous \galex\ match, the UV
measurement may still be affected by blending of several objects
independently detected in SDSS. We will address this issue and
describe the implemented solution in Section \ref{sec:data}.

\begin{figure}
\epsscale{1.1} \plotone{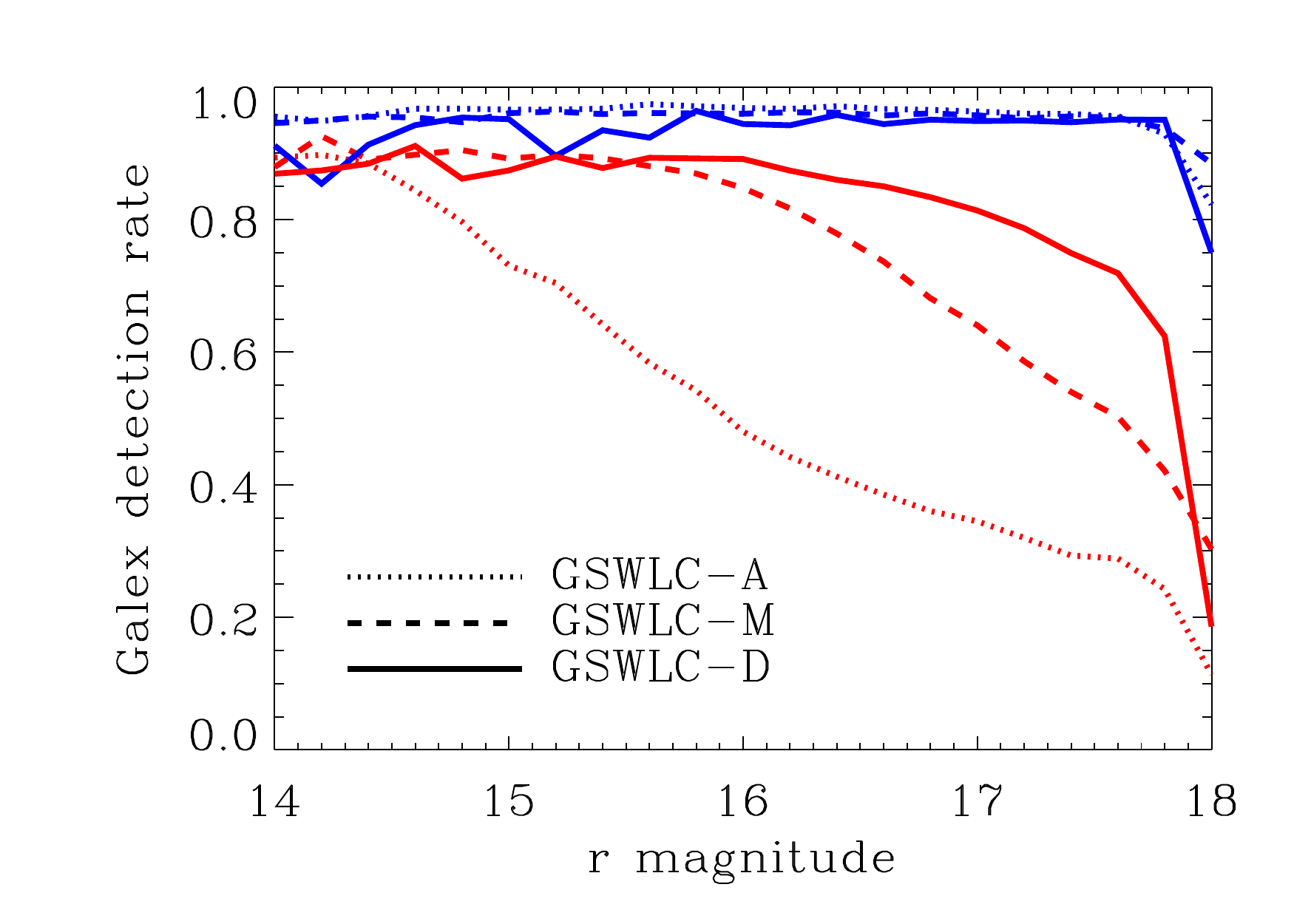}
\caption{UV detection rates of different GSWLC catalogs (defined by UV
  survey depth). Detection rate is shown as a function of $r$
  magnitude, separately for blue ($g-r<0.7$, observer frame; blue
  lines) and red galaxies (red lines). The detection rate of red
  galaxies is strongly dependent on UV depth and
  magnitude.\label{fig:glx_det}}
\end{figure}

UV detection rates of SDSS objects ($3\sigma$ threshold, either UV
band) are 54\% for GSWLC-A, 74\% for GSWLC-M and 84\% for
GSWLC-D. Besides the UV survey depth, the detection rates depend
strongly on galaxy's SSFR or color, as can be seen from Figure
\ref{fig:glx_det}, where we plot the detection rates as a function of
$r$-band magnitude, separately for blue ($g-r<0.7$) and red galaxies.

\subsection{SDSS-\wise\ and 2MASS matching}\label{ssec:match2}

GSWLC utilizes \wise\ observations at 22 $\mic$ (\wise\ channel W4) to
determine SFRs independently from the UV/optical SED fitting. The
depth of \wise\ observations over the sky is not uniform ($6\pm1$ mag
Vega), but is still much more uniform than \galex\ depth, and
essentially covers the entire sky without gaps. FWHM of W4 PSF is
$12\arcsec$. We use 22 $\mic$ photometry from two independent
reductions of \wise\ data. The first is the official AllWISE Source
Catalog\footnote{\url{http://wise2.ipac.caltech.edu/docs/release/allwise/}.},
hosted at IPAC Infrared Science Archive (IRSA) at the Caltech/JPL, and
the second is the ``unWISE'' reduction
\citep{lang16}\footnote{\url{unwise.me}.}, where SDSS detections
served as forced photometry priors.

To match SDSS to AllWISE, we take the closest candidate in a
$5\arcsec$ search radius. Note that AllWISE detections and astrometry
are based on simultaneous PSF-matched fit to all channels. Thus, the
shorter-wavelength channels (3.4 (W1) and 4.6 $\mic$ (W2)), which have
smaller PSF widths ($6 \arcsec$) and higher S/N, will dominate the
astrometric accuracy ($<0\farcs5$). This makes the adopted search
radius sufficient to capture genuine matches, and justifies taking the
closer candidate in case of multiple matches (3.2\% of cases). The
detection rate of SDSS targets at 22 $\mic$ (based on $2\sigma$
threshold of profile magnitudes) is 32\%, which is lower than the
detection rate of \galex.

Since unWISE catalog is derived based on SDSS DR10 detections, no
matching is required (sources already carry SDSS {\tt ObjID}). Forced
photometry increases the S/N of measured fluxes, so the $2\sigma$
detection rate is 41\%,a third higher than of the official catalog.

Near-IR photometry ($JHK_s$) is taken from 2MASS Extended Source
Catalog (XSC), also hosted at IRSA. We match SDSS to 2MASS XSC by
selecting the closest candidate within $5\arcsec$ search radius. The
detection rate of SDSS targets is 48\%.

\section{Data}\label{sec:data}

The data used to construct GSWLC consist of broad-band photometry
(magnitudes or fluxes and their errors) from \galex, SDSS and \wise,
and redshifts from SDSS. Comparison datasets will also use line fluxes
from SDSS spectra.

\subsection{Optical photometry}

SDSS offers several choices for galaxy photometry: {\tt modelMag}
(magnitude extracted assuming either de Vaucouleurs or exponential
profile), {\tt cmodelMag} (weighted average of de Vaucouleurs and
exponential magnitudes) and {\tt petroMag} (surface-brightness
dependent aperture magnitude). The usual practice in SED fitting is to
use fluxes that yield most accurate colors. Of SDSS magnitudes, the
{\tt ModelMag} best fulfills that role. We confirm that this is the
optimal choice by finding that the best-fitting models (based on
optical fitting alone) have three times lower median $\chi^2$ when
{\tt modelMag} are used as opposed to {\tt cmodelMag} or {\tt
  petroMag}. Also, SSFRs obtained from UV/optical SED fitting that
uses {\tt modelMag} magnitudes have a smaller scatter with respect to
SSFRs from \wise\ and from \ha. While {\tt ModelMag} produces stable
colors, the degree to which it will estimate the total light will
depend on galaxy size, morphology \citep{bernardi10,taylor11} and
color gradient. \citet{simard11,meert16} have derived magnitudes that
were designed to better capture the total light in each band. Readers
can use those catalogs to scale our estimates of the stellar mass and
SFR, if that is required for their goals. Based on the comparison of {\tt
  ModelMag} and \citet{meert16} $r$-band magnitude, the typical
correction should nevertheless be relatively modest ($+0.03\pm0.08$
dex).

\subsection{UV photometry and corrections}

The \galex\ source catalog offers only one measure of flux that is
recommended for galaxies: {\tt MAG\_AUTO}, a Kron elliptical aperture
magnitude derived by SExtractor \citep{bertin}.  We investigate the
differences in the methodology of this measurement compared to the one
used for SDSS photometry, as well as the effects of lower \galex\
resolution, by deriving {\tt MAG\_AUTO} magnitudes of 1000 randomly
selected SDSS targets whose $g$-band images were degraded to match
\galex\ resolution. By comparison with isolated sources, we find that
the differences between \galex-resolution {\tt MAG\_AUTO} and SDSS
resolution {\tt modelMag} are dominated by systematic offsets (up to
0.4 mag) arising from close companions (especially at $d<10\arcsec$)
which become blended in lower resolution, rather than the differences
in the methodology of magnitude measurement ($<0.1$ mag difference).

To account for the systematics that arise from both the difference in
\galex\ and SDSS resolutions and the methods of flux measurement, we
derive and implement several corrections to UV fluxes. Corrections are
derived by comparing the UV magnitudes predicted from optical-only SED
fitting of star-forming galaxies ($g-r<0.7$) with the actual UV
magnitudes. While any individual predicted UV magnitude is crude
($\sigma \approx 0.4$ mag in NUV), the corrections are accurate
owing to a very large number of galaxies that define them.

First, we confirm that there are no zero-point offsets between the
predicted and the actual UV magnitudes.  Next, we find that a small
edge-of-detector correction is required for NUV, but not FUV,
photometry.  NUV magnitudes of sources that appear in the outermost
$8'$ annulus of the field of view, which accounts for 40\% of the
detector area, are up to 0.1 mag too bright, possibly because of an
inaccurate flat fielding. The amount of correction depends linearly on
the radial position within the annulus:

\begin{equation}
{\rm NUV}_{\rm corr} = {\rm NUV} + 0.78\: {\rm FOV}-0.37,\quad {\rm FOV}>0.47
\end{equation}

\noindent where FOV is the distance from the center of the tile in
degrees.

Next, we derive the centroid shift correction. The shift between
optical and UV positions arises due to random errors, mostly from the
lower accuracy of the \galex\ astrometry, but also from the
differences in UV versus optical morphology, which has a slight effect
on the measured magnitudes. Again, using the difference between
predicted and observed NUV magnitudes, we find the following
correction:

\begin{equation}
{\rm UV}_{\rm corr} = {\rm UV} + 0.054\; {\Delta x}-0.049,\qquad
\Delta x>0\farcs7
\end{equation}

\noindent where $\Delta x$ is the shift between \galex\ and SDSS
positions in arcsec. Correction is not applied when $\Delta x\leq
0\farcs7$, where random errors dominate. We apply the same correction
to NUV and FUV magnitudes.

By far the most significant correction for UV magnitudes is due to
blending. The degree to which a \galex\ magnitude will be affected by
blending will depend both on the relative brightness of the nearest
companion (any SDSS photometric object) and its separation ($d$) from
the target. We construct corrections (same for NUV and FUV) as a
function of the difference in $g$ magnitude ($\Delta g = g_{\rm
  comp}-g_{\rm target}$) for four ranges of separation:

\begin{equation}
\begin{array}{l}
{\rm UV}_{\rm corr} = {\rm UV} -0.036\: \Delta g + 0.37, \qquad d<4\arcsec \\ 
{\rm UV}_{\rm corr} = {\rm UV} -0.046\: \Delta g + 0.35, \qquad
4\arcsec\leq d<10\arcsec \\
{\rm UV}_{\rm corr} = {\rm UV} -0.019\: \Delta g + 0.15, \qquad
10\arcsec\leq d<15\arcsec \\
{\rm UV}_{\rm corr} = {\rm UV} -0.006\: \Delta g + 0.04, \qquad
15\arcsec\leq d<20\arcsec 
\end{array}
\end{equation}

We note that even though the corrections to UV magnitudes may be
significant for individual objects (in particular those with a close
blended companion), the SFRs for the majority of star-forming galaxies
are not significantly affected by their application.

We use FUV fluxes measured at NUV positions ({\tt FUV\_NCAT}) rather
than from independent FUV detections, as they provide more robust UV
color. Both NUV and FUV measurements require a $3\sigma$ threshold in
{\tt FLUX\_AUTO}.

\subsection{Galactic reddening and calibration errors}

SDSS and \galex\ photometry must be corrected for galactic
reddening. We find that the combination of extinction coefficients
from \citet{peek13} for UV bands and \citet{yuan13} for optical bands
produces somewhat better fits ($\sim\! 10\%$ smaller $\chi^2$) than the
corrections based on \citet{sfd}, or on \citet{yuan13}
alone. We reproduce the adopted corrections here.

\begin{equation}
\begin{array}{c}
A_{\rm FUV} = 10.47\, E(B-V)+8.59 \, E(B-V)^2-82.8 \, E(B-V)^3 \\
A_{\rm NUV} = 8.36\, E(B-V)+14.3 \, E(B-V)^2-82.8 \, E(B-V)^3 \\
A_u = 4.39 \, E(B-V) \\
A_g = 3.30 \, E(B-V) \\
A_r = 2.31 \, E(B-V) \\
A_i = 1.71 \, E(B-V) \\
A_z = 1.29 \, E(B-V)
\end{array}
\end{equation}

Photometry catalogs usually list only random flux errors, without
systematic, or calibration errors. We find that the default CIGALE
padding of formal photometry errors by the addition (in quadrature) of
0.1 mag to catalog magnitude errors to account for systematic errors
in models and photometry leads to a significant loss of derived
physical parameter accuracy, as evidenced by 50\% larger scatter when
derived SFRs are compared with independent SFR estimates.  Instead,
following S07, we add more modest calibration errors (determined from
repeat observations) of (0.052, 0.026, 0.02, 0.01, 0.01, 0.01, 0.01)
mag in (FUV, NUV, $u$, $g$, $r$, $i$, $z$), plus the $u$-band red leak
error of $\sigma_{u, {\rm RL}} = 0.0865(r-i)^2+0.0679(r-i)$, derived
in S07 based on the description of red leak in \citealt{sdssdr2}.  The
validity of the adopted calibration errors is verified by constructing
the distribution of magnitude residuals (fitted minus real magnitude
divided by the total error), which follow unit Gaussians.

\subsection{Mid-IR photometry and ancillary data}

We derive what we refer to as ``mid-IR SFRs'' using two types of 22
$\mic$ magnitude (flux) measurements. The first is what AllWISE Source
Catalog calls a {\it profile-fitting} magnitude ({\tt w4mpro}). The
profile is simply the PSF, so these magnitudes are essentially PSF
magnitudes, and are thus most appropriate for unresolved
sources. Since the PSF FWHM at 22 $\mic$ is $12\arcsec$ , which is
larger than 91\% of galaxies ($r_{90}$ size), these magnitudes are
reasonably appropriate for SDSS galaxies. Yet, PSF magnitudes will
systematically underestimate the flux in larger galaxies. This is
largely remedied in forced photometry from unWISE catalog, which
applies SDSS-measured galaxy profiles (convoluted with W4 PSF) as
photometry priors.

To derive emission-line SFRs (which we call ``\ha\ SFRs''), we use
\ha\ and \hb\ fluxes from the MPA/JHU
catalog, based on DR7\footnote{\url{http://www.mpa-garching.mpg.de/SDSS/DR7}\\
  \url{http://home.strw.leidenuniv.nl/~jarle/SDSS/}.}. Measurement of
emission line fluxes is described in \citet{tremonti04}. We also use
the emission-line (BPT diagram, \citealt{bpt}) classification of
galaxies from the MPA/JHU catalog, derived as described in \citet{b04}
(hereafter B04).

\section{Derivation of mid-IR and emission-line SFRs} \label{sec:wise}

Although CIGALE allows the IR dust emission to be fit in conjunction
with the stellar emission, we derive the mid-IR SFRs separately, to
avoid potential systematics.  For example, CIGALE normalizes the IR
SED template, so that the total dust emission (i.e., total IR
luminosity, $L_{\rm IR}$) equals the stellar emission absorbed by the
dust in the UV/optical/near-IR. However, the shape of the IR SED is a
free parameter \cite{noll09}, which means that without the far-IR SED,
as in the case of \wise\ data, the IR luminosity will not be strongly
constrained, and could thus be driven, through energy balance
requirement, by any potential systematics in the estimate of the
absorbed stellar luminosity, such as those that would arise from
incorrect assumptions about the attenuation law.

Instead, accurate IR luminosities can be obtained from 22 $\mic$
observations using {\it luminosity-dependent} IR templates, and these
estimates can then be compared to the results of the SED fitting of
the stellar emission (for actively star-forming galaxies, where
dust heating by young stars dominates), to verify whether energy
balance is satisfied.

Another advantage of performing the stellar SED fitting separately is
that in a joint fit to stellar and dust emission one also has to worry
about the contamination of IR emission from dust-obscured Type 2 AGN
(such as Seyfert 2, or high-excitation radio galaxies,
\citealt{yan13,pace16}). While the AGN contribution can be modeled in
CIGALE, it introduces additional degrees of freedom, further weakening
the usefulness of an energy balance requirement, in particular when
only mid-IR photometry is available.

The mid-IR SFRs that we report in GSWLC are computed as follows.
First we calculate the total IR luminosity (8-1000 $\mic$) by
interpolating the luminosity-dependent IR templates of \citet{ce01} so
that they match the 22 $\mic$ flux.  We also tested using \citet{dh02}
templates, which do not have an associated luminosity, but the IR SED
shape-luminosity dependence is imposed from empirically-calibrated
relations of \citet{marcillac06}. The two methods produce very similar
IR luminosities (average difference 0.01 dex, scatter 0.02 dex), but
\citet{ce01}-based luminosities are marginally ($<2\%$) better
correlated with IR luminosities from {\it Herschel}, so we adopt them
for GSWLC.

\begin{figure}
\epsscale{1.0} \plotone{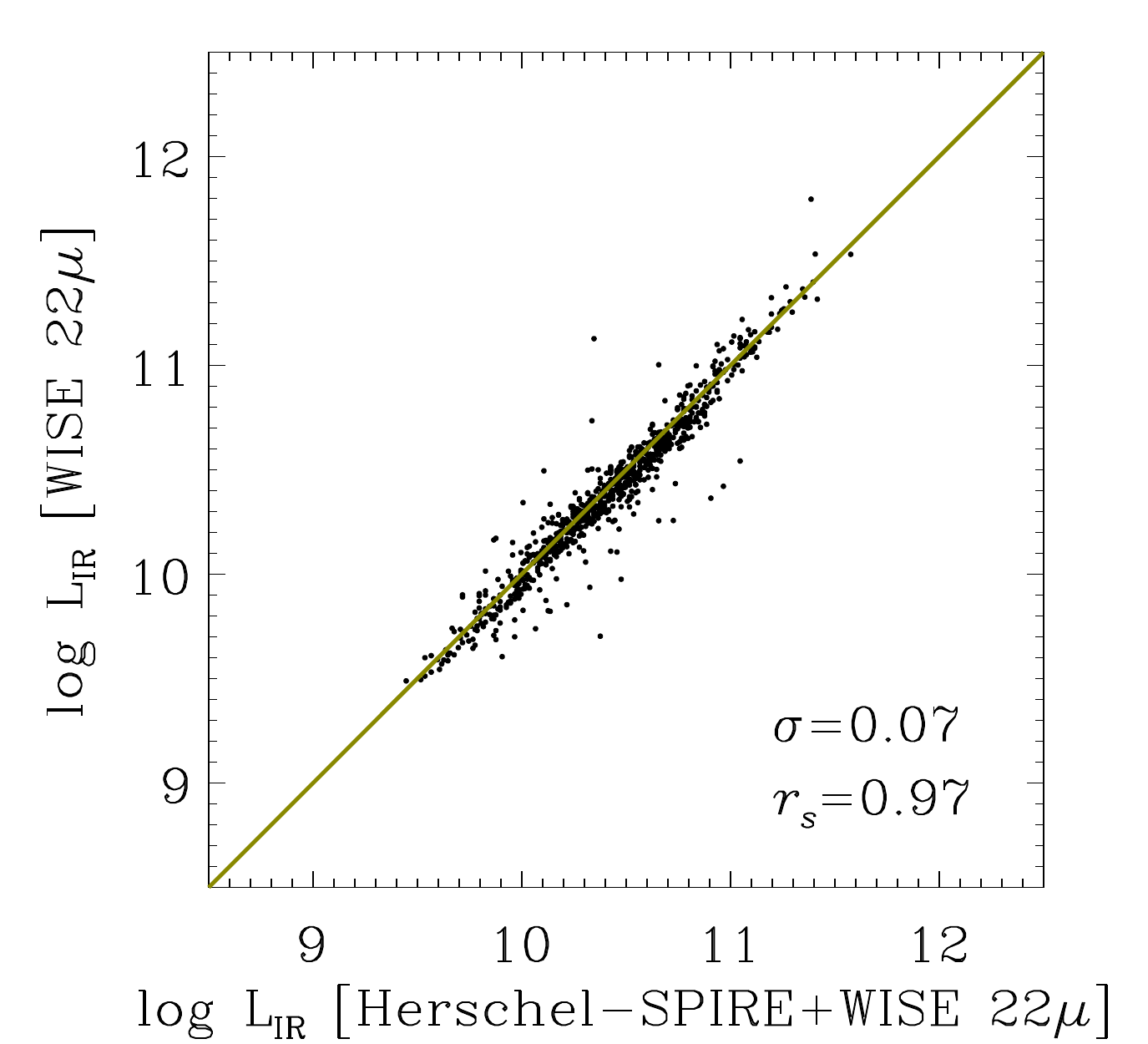}
\caption{Comparison of IR luminosities derived from \wise\ 22 $\mic$
  flux, against {\it Herschel+WISE} IR luminosities from
  \citet{rosario16}. Our \wise\ IR luminosity is based on a single
  flux point in the mid-IR, converted to total IR luminosity using the
  luminosity-dependent IR templates of \citet{ce01}, whereas the {\it
    Herschel+WISE} IR luminosity comes from 22 $\mic$ and sub-mm 
  (250--500 $\mic$) measurements. The two estimates agree very well. Solid line is
  a 1:1 relation.
  \label{fig:herschel}}
\end{figure}

We check the robustness of our 22 $\mic$-derived IR luminosities by
comparing them to IR luminosities derived for SDSS Stripe82 galaxies
by \citet{rosario16}. Their IR luminosity is obtained by fitting IR
templates to \wise\ 22 $mic$ flux {\it and} the sub-mm fluxes from
{\it Herschel} SPIRE (250, 350 and 500 $\mic$). Because they have
multiple flux points, they derive IR luminosities from \citet{dh02} IR
templates that do not have an associated luminosity. The cross-matched
sample consists of 817 galaxies. As can be seen in Figure
\ref{fig:herschel}, the agreement of IR luminosities is excellent,
with no ($<0.01$ dex) systematic offset over the range probed by {\it
  Herschel} ($9.5<\log L_{\rm IR}<11.4$), perfect linearity, and a
scatter of only 0.07 dex. Comparison of the specific luminosities
($L_{\rm IR}/M_*$), which removes the distance dependence (Section
\ref{sec:comp}), reveals a mild non-linearity (with a slope of 1.10),
in the sense that \wise\ IR luminosities of galaxies with the lowest
specific luminosity (or SSFR) are 0.1 dex lower compared to galaxies
with the highest specific luminosity. While the agreement between our
and Rosario et al.\ IR luminosities is encouraging, there is a caveat
that the latter are not fully independent form our estimates, because
both use 22 $\mic$ flux. The role of 22 $\mic$ flux point in Rosario
et al.\ estimate is likely significant given that their data do not
sample the peak of the IR SED ($\sim 100 \mic$).

The above comparison was carried out for 22 $\mic$ IR luminosities
derived using profile (PSF) fluxes from the official AllWISE
catalog. If we instead used 22 $\mic$ photometry from unWISE
\citep{lang16}, the small scatter with respect to \citet{rosario16}
{\it Herschel+WISE} IR luminosities of $\sigma=0.07$ dex (Figure
\ref{fig:herschel}) would grow to $\sigma=0.14$. This does not
necessarily imply that unWISE fluxes are noisier, because the {\it
  Herschel+WISE} uses the PSF flux from AllWISE catalog.

To obtain mid-IR SFRs from IR luminosity, we use a simple conversion
given by \citep{kennicutt98}, adjusted to Chabrier IMF using the 1.58
conversion factor (S07):

\begin{equation}
\log {\rm SFR} = \log L_{\rm IR} - 9.966
\label{eq:lir}
\end{equation} 

\noindent where SFR is in $\msperyr$ and $L_{\rm IR}$ in
$L_{\odot}$. 

To derive \ha\ SFRs, which we use for comparison purposes, we take
\ha\ fluxes and correct them using the Balmer decrement method,
assuming Case B recombination and the \citet{cardelli89} extinction
curve. Emission line fluxes are measured within $3\arcsec$ SDSS
fibers, which we find to capture, on average, 30\% of the galaxy mass,
depending on galaxy's distance, size and profile.  Dust-corrected \ha\
fluxes are then converted to SFR using the \citet{kennicutt98}
relation, adjusted by $-$0.24 dex \citep{muzzin10} to convert from
Salpeter to Chabrier IMF.  Comparison of fiber \ha\ SFR with other,
total SFR indicators, can be accomplished if all measurements are
converted to SSFRs.  The \ha\ SSFR is obtained by normalizing by the
stellar mass present within the fiber, which are taken from the
MPA/JHU catalog.  \footnote{Stellar masses and SFRs from the MPA/JHU
  catalog have been adjusted from Kroupa to Chabrier IMF by applying
  $-$0.025 dex offset (S07).}

\section{Derivation of physical parameters from UV/optical SED
  fitting} \label{sec:sed}

\subsection{Methodology}

Physical parameters from UV/optical SEDs are derived using CIGALE
(\citealt{noll09}, M.\ Boquien et al.\, in prep.), a powerful code
that produces libraries (grids) of model SEDs and performs the SED
fitting. The code was originally written in Fortran and recently
re-written in Python. It is continually being improved and expanded in
capabilities. We use the most recent Python version 0.9, which offers
full control over parameters that specify SF history, dust
attenuation, and emission line fluxes.  This flexibility has proved
essential in order to derive robust results.

Model SEDs produced by CIGALE, and from which model photometry is
extracted, consist of UV/optical/near-IR stellar emission ($\lambda
\lesssim 5 \mic$), and, optionally, the IR dust emission ($\lambda
\gtrsim 5 \mic$). Third, also optional component is the nebular
emission (lines and continuum). For reasons discussed in Section
\ref{sec:wise}, we will restrict the SED fitting to stellar emission,
with the contribution of emission lines included.

CIGALE synthesizes stellar emission based on either \citet{bc03}
(BC03) or \citet{m05} (M05) SPS. The most important difference between
them is in the treatment of the TP-AGB phase of stellar evolution
\citep{maraston11,marigo15}. CIGALE cannot currently include the
contribution of emission lines in conjunction with M05 models, which,
as we will demonstrate, are important for obtaining robust
measurements. Therefore we opt to use BC03 models. The use of BC03
models may anyhow be a more appropriate choice, given some evidence
that BC03 models better reproduce optical and near-IR colors than M05
models \citep{kriek10,conroy10a,zibetti13}.

CIGALE can also perform Bayesian SED fitting and report the physical
parameters and their errors. Bayesian SED fitting consist of building
probability distribution functions of physical parameters by assigning
probabilities to each model spectrum (at a matching redshift) based on
the goodness of fit between the model and observed broad-band SEDs.
The methodology has been described in detail in S07, \citet{dacunha08}
and other papers.  \cite{walcher11} and \citet{conroy13} provide
comprehensive reviews of various aspects of SED fitting. While SED
fitting represents an optimal way to extract information from
photometry (and/or spectra), it will be limited by the uncertainties
in stellar and dust attenuation models
\citep{conroy09,conroy10a,conroy10b,mitchell13}, and the choice of
priors (e.g., appropriate SF histories and dust attenuation laws).

In order to specify the modeling assumptions (e.g., which dust model
or parameterization of SF history to use) the usual practice is to
select the parameters that yield the smallest relative differences
between the observed and the best-fitting model photometry, i.e., that
minimize some average $\chi^2$ of the sample. However, in some cases
different modeling assumptions may lead to marginal changes in
$\chi^2$ (suggesting that UV/optical SED is degenerate with respect to
them). We will therefore require both the internal consistency (small
$\chi^2$) and the external one: maximum correlation of the SED fitting
SFR with respect to two entirely independently derived SFR tracers
(mid-IR and \ha).

In our analysis we assume a Chabrier IMF \citep{chabrier}.  CIGALE
assumes a flat WMAP7 cosmology ($H_0=70$ km s$^{-1}$ Mpc$^{-1}$,
$\Omega_m=0.27$).

\subsection{Near-IR photometry}

We perform SED fitting on UV and optical photometry, (0.15-0.9
$\mic$), omitting the longer-wavelength stellar emission (1.1-5.4
$\mic$).  The near-IR photometry is excluded because it increases the
reduced $\chi^2$ values and worsens (albeit slightly) the correlations
with mid-IR and \ha-derived SFRs.  Parameters from the SED fitting are
reliable to the extent that the models are able to reproduce the
observed colors. We find that the model photometry, either based on
BC03 or M05, does not accurately reproduce the near-IR colors from
2MASS photometry. In particular, both models imply a strong
correlation between $J-K$ and $i-J$, which is not seen in the data
(S.\ Salim, in prep.)

Stellar mass estimates based on near-IR photometry are often perceived
as more accurate and/or more precise than those based on optical
bands, or that, at the minimum, the near-IR photometry improves the
precision of stellar mass estimates.  The reasoning underlying this
claim is that stellar emission peaks in the near-IR and that near-IR
mass-to-light ratio is less sensitive to the stellar population age,
and the effects of dust. While these arguments are correct, they do
not take into account large modeling discrepancies in the near-IR
mass-to-light ratios \citep{mcgaugh14}, arising due to the
uncertainties in our understanding of the post-main sequence phases of
stellar evolution, the phases which dominate the energy output in the
near-IR \citep{conroy13}. For example, \citet{vanderwel06} find that
the inclusion of near-IR photometry in SED fits leads to discrepancies
with respect to dynamical masses, and that these discrepancies depend
on the SPS model used.

Even if there existed no major uncertainties in the models and no
discrepancies between the models and the data (thus, no concerns that
the near-IR will affect the {\it accuracy} of mass estimates), the
improvement in the {\it precision} of stellar masses achieved by
adding the near-IR photometry to the optical photometry is much more
modest than usually assumed. \citet{taylor11} have studied physical
parameters derived from mock observations of GAMA galaxies when just
the optical photometry, or optical plus UKIDSS-depth near-IR
photometry was used to perform the SED fitting. They show that the
improvement in the precision of stellar masses when the near-IR
photometry is added to SED fitting is small (0.05 vs.\ 0.06 dex, their
Fig.\ A2). The parameter whose precision improves the most is the
stellar metallicity.  Indeed, the sensitivity of near-IR luminosities,
and even red optical bands ($i$ or $z$), to metallicity (e.g., Fig.\ 9
of \citealt{courteau14}; Fig.\ 10 of \citealt{taylor11}) is another
reason why near-IR alone is a more problematic tracer of stellar mass
than the multi-band optical light.

\subsection{Star formation histories}

In the following three sections we discuss the choice of modeling
parameters, starting with SF histories and metallicities.

BC03 models are available for six stellar metallicities, of which we
use the higher four, from 0.2 $Z_{\odot}$ to 2.5 $Z_{\odot}$, which is
an adequate range for most galaxies in SDSS \citep{gallazzi05}.

We have considered two different parametrizations of SF history
offered by CIGALE: two-component exponential and delayed
exponential. In the two-component exponential model, the SF history is
a composite of two exponentially declining functions ($\tau$ models),
each with its own starting epoch (age) and e-folding time. The
normalization of the younger population is specified by the mass
fraction $f$, which can also be set to zero, resulting in a single
exponential model. Each component starts at maximum value and then
decreases monotonically. In contrast, the delayed exponential SF
history is smooth (e.g., \citet{gavazzi02}): it starts from zero SFR,
reaches a peak at some time and then declines.  It is given by:

\begin{equation}
{\rm SFR} \propto \frac{t}{\tau^2} e^{-t/\tau} \label{eq:delayed}
\end{equation}

We find that as long as the parameters are chosen so that the model
colors (and more specifically, model SSFRs) cover the range of colors
or SSFRs present in observations (for delayed SFHs this requires
allowing $\tau<0$ in Eq.\ \ref{eq:delayed}), either SF history will
produce similar stellar masses and SFRs. Specifically, there is no
systematic difference for masses of non-main sequence, low-SSFR
galaxies (log SSFR$<-11$), while the difference in masses for actively
star-forming galaxies is typically 0.1 dex. The difference in SFRs is
similar in degree (see also \citealt{boquien14}), and of the same
sign, which means that the difference in SSFRs is less than either the
difference in mass or SFR. While both parametrizations perform
reasonably well, the two-component exponential SF histories yield
better fits (geometric mean of reduced $\chi^2$ 0.7 vs.\ 1.0),
presumably because, by not being smooth, they are able to better match
the bursty SF history of low-mass galaxies \citet{weisz11}. We thus
adopt the two-component exponential parametrization.

Delayed exponential models have been preferred over exponential models
in some recent studies (e.g., \citealt{simha14}). However, there are
important differences among the exponential models. If a single
exponential model is assumed to have started in the early universe
(e.g., $t_0\sim 10^{10}$ yr), in order to acknowledge the fact that
galaxies contain ancient stellar populations no matter how dominant
the current episode of SF may be \citep{aloisi07}, then such model
will fail to reproduce the SSFRs of many star-forming galaxies today,
simply because log SSFR$_{\rm max} = \log t_0^{-1} = -10$. It is this
naive implementation of the exponential model (e.g.,
\citealt{simha14}) that results in inferior performance. The problem
of single exponential model not producing high SSFRs can be alleviated
by allowing the starting epoch of the model to be more recent than the
Big Bang. Though later epoch galaxy {\it formation} is obviously not
realistic, it can instead be interpreted as the epoch of peak of SF
activity. Variable starting time was assumed in essentially all work
that base their SF histories on a single exponential, including S07,
which in addition had stochastic bursts superimposed on exponential
models.

Allowing recent starting times for the exponential model results in
SFRs that are comparable to more sophisticated
models.\footnote{\citet{pacifici15} report large (0.6 dex, on average)
  offsets at $z>1$ between SFRs derived using their classical models
  (which assume exponential histories with a range of formation times)
  and their more sophisticated models (SF history based on
  cosmological simulations). However, the two models also assume
  different attenuation laws (fixed slope vs.\ multiple slopes), which
  will affect the SFRs through the derived dust attenuation (see their
  Figure 6.), so it is not clear that the difference in SF histories
  is responsible for most of the offset.}  However, this approach will
neglect (``outshine'') any old population, resulting in somewhat
underestimated masses \citep{papovich01,michalowski14}. This problem
is resolved with a two-component exponential model used here, where
one component corresponds to high-redshift SF.  While more realistic
SF histories may have multiple bursts, or are in general quite
variable \citep{weisz11}, modeling these features with high temporal
precision is unimportant from the standpoint of broad-band SED
fitting, where even the characteristics of the most recent burst
cannot be constrained with any precision because of the burst age vs.\
burst amplitude degeneracy (e.g., \citealt{smith15}).

The parameters of the adopted two-component exponential model are as
follows.  The formation time of the old population is 10 Gyr before
observation epoch, with e-folding times that span a range from 850 Myr
(fast decline that defines the lower SSFR limit of log SSFR $=-13.8$)
to nearly constant 20 Gyr (which has declined only 0.2 dex since
formation). Formation times for the younger component span from 100
Myr (the shortest timescale to which UV observations are sensitive) to
5 Gyr. Their e-folding time is 20 Gyr, i.e., nearly constant.
In order to produce the highest SSFRs observed today (log SSFR
$\sim\!-8$), the mass fraction of the younger component must span up
to $f=0.5$.  Thus our library of SF histories looks like old
exponentials with various decay times, with a relatively flat burst
superimposed.

\begin{figure*}
\epsscale{0.66} \plotone{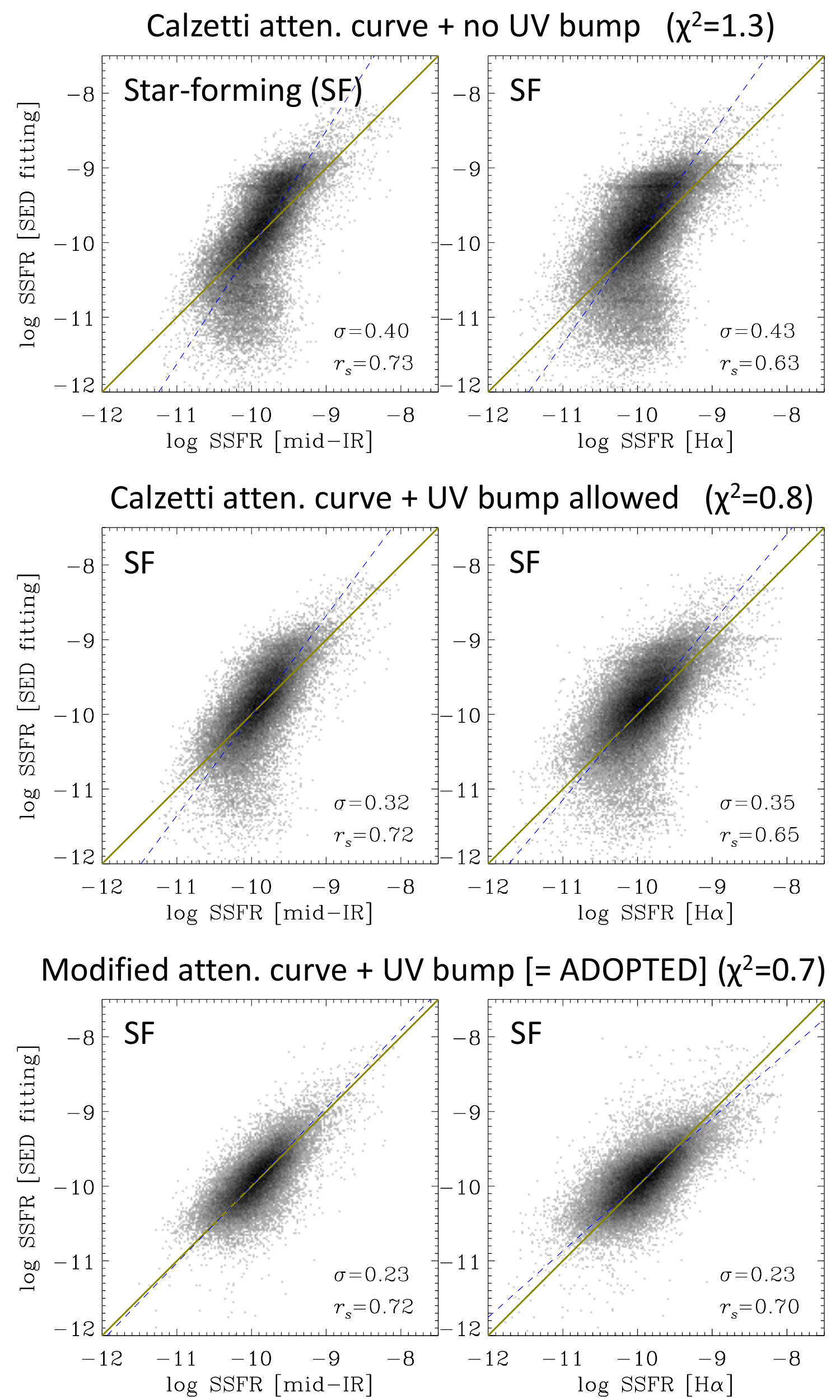}
\caption{The choice of attenuation law to use in UV/optical SED
  fitting. We show comparison between specific SFRs (SSFR) obtained
  from the UV/optical SED fitting (assuming different attenuation
  laws) and SSFRs based on \wise\ 22 $\mic$ observations (mid-IR SSFR;
  left panels) and from Balmer-decrement corrected \ha\ (right
  panels). The upper panels use the standard Calzetti attenuation
  curve. Strong trends against both SFR tracers are seen. They are
  reduced by allowing for a UV bump (the middle row), which also
  improves the formal $\chi^2$ of SED fits. Further improvement,
  resulting in very good match with mid-IR and \ha\ SSFRs, is achieved
  by also making the attenuation curve steeper than the Calzetti one
  (bottom row).  The changes in SSFRs are mostly driven by the changes
  in SFRs, rather than the stellar masses. Full line is a 1:1
  relation, and the dashed line is the robust bisector linear fit
  (used throughout the paper). Standard deviation around the fit, as
  well as the Spearman correlation coefficient, are given in each
  panel. For these and subsequent figures (except where noted) the
  lower redshift bound is 0.025 in order to reduce AllWISE photometry
  systematics for galaxies having large angular size. In order to
  present a large number of data points without blotting, we use
  grayscale where the shade scales as number of galaxies in pixel to
  the power of 0.3. Sublinear exponent is chosen to better show the
  outliers. Shown are the data from GSWLC-M, but similar results are
  obtained with GSWLC-A or D. \label{fig:ssfr_comp}}
\end{figure*}

\begin{figure*}
\epsscale{0.66} \plotone{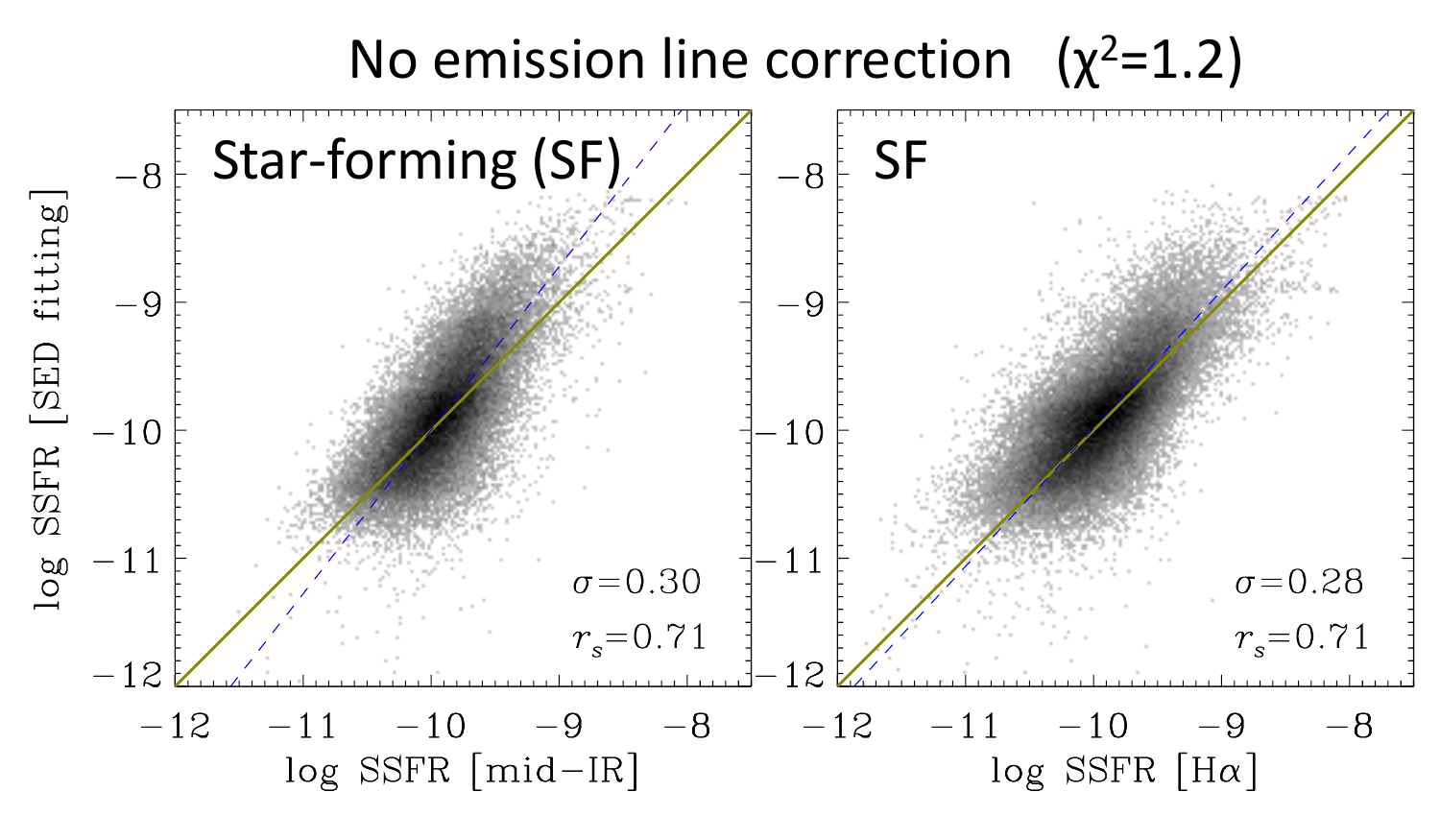}
\caption{The effect on SSFRs of not correcting for nebular emission
  lines. This figure should be compared to the lower panels of Figure
  \ref{fig:ssfr_comp}, where otherwise identical models include the
  contribution of emission lines. Not correcting for emission lines
  produces overestimates of SED (S)SFRs of high-SSFR galaxies, greater
  random errors (quantified as the standard deviation around the fit,
  blue dashed line) and also the increased mean reduced
  $\chi^2$. \label{fig:ssfr_nolines}}
\end{figure*}

\subsection{Dust attenuation laws}

CIGALE allows the dust attenuation law to be specified as a
simple power law ($k(\lambda)_{\rm PL} \propto \lambda^{\alpha}$,
following \citealt{cf00}), or according to the \citet{calzetti00} recipe
($k(\lambda)_{\rm Cal}$). The Calzetti curve can be
modified so that the overall slope becomes more or less steep, by
multiplying it with the power law of slope $\delta$ \citep{noll09}:

$$
k(\lambda)_{\rm mod} = k(\lambda)_{\rm Cal} (\lambda/5500\text{\AA})^{\delta}
$$

\noindent Note that slope modification changes $R_V$ from 4.05, the
value it has for the Calzetti curve. The attenuation law can be
further modified to include a UV bump of varying strength
\citep{stecher65,fitzpatrick86,conroy10uv}.

We systematically tested various modifications of the Calzetti
attenuation law. We produce model grids where the range of color
indices lies between $E(B-V)=0.1$ and 0.6 mag for the young ($<10$
Myr) population (i.e., the dust affecting the nebular lines;
equivalent to $A_{\text \ha}=0.25$--1.5 mag).  Attenuation that
affects older stars, and therefore most of the stellar continuum, is
taken to be smaller by a factor of 0.44, following
\citet{calzetti00}. The tests were carried out with galaxies
classified in the BPT diagram as star-forming, for which mid-IR
luminosity and dust-corrected \ha\ luminosity provide reliable
estimates of SFR.  The results are illustrated in Figure
\ref{fig:ssfr_comp}. The upper panels show the comparison between SSFR
obtained from the SED fitting, assuming the standard Calzetti curve,
which has no UV bump, against independently derived mid-IR and \ha\
SSFRs.  Note that the comparison is carried out in terms of SSFR,
which better reveals systematic trends than the usual SFR comparison
(Section \ref{sec:comp}).  Similar trends are seen vs.\ mid-IR and
\ha.  The differences increase with the increasing SSFR until log
SSFR$_{\text SED}=-9$, in the sense that the values derived from SED
fitting assuming the Calzetti curve tend to be up to 0.4 dex higher.
At still higher SSFRs however, there is reasonable agreement between
the SSFRs, as expected, since the Calzetti curve was derived from
starburst galaxies with high SSFRs.

The middle row shows the comparisons when the Calzetti law is modified
to include a UV bump of varying intensity (from no bump to $4\times$
the MW value). The UV/optical SED fits become formally better
(geometric mean of reduced $\chi^2$s goes from 1.3 to 0.8), and the
correlation with mid-IR and \ha\ SSFRs improves (from $\sigma=0.40$ to
0.32 dex, for mid-IR comparison). However, the non-linearity (the
slope of the correlation featuring logarithms of SSFRs not being
unity) persists, as well as a tail of galaxies with unusually low
SSFRs.

The bottom row shows the comparisons when the Calzetti attenuation law
is further modified to make it steeper (with $\delta = -0.5$ and
$-1.0$), in addition to allowing the UV bump. There is now a
significant reduction in scatter with respect to mid-IR and \ha\ SSFRs
($\sigma=0.23$), and the relation is fairly linear. This is our
adopted dust attenuation model.  We note that achieving this level of
agreement between SED fitting SSFRs and mid-IR SSFRs requires both the
steepening of the attenuation curve and the addition of the UV
bump. Just steepening the slope of the attenuation curve does remove
most of the non-linearity with respect to mid-IR SSFR (plot not
shown), but still yields relatively high reduced $\chi^2$ of 1.0,
regardless of the amount of steepening. In other words, the
combination of the two modifications is necessary to achieve both the
small $\chi^2$ values and good agreement with other indicators.

The agreement between SED and mid-IR SFRs demonstrates that the energy
balance is fulfilled when this modified attenuation law is used: the
energy absorbed by the dust in the stellar SED matches the energy
emitted in the IR.

We note that assuming an attenuation curve in a power-law form, having
slopes $\alpha=-1.0$ (preferred by the majority of galaxies) and
$\alpha=-1.5$, and adding the UV bump to them, has a similar effect to
modifying the standard Calzetti curve as described in the preceding
paragraph.  For comparison, the standard Calzetti curve can be
approximated in the UV range by a power-law of slope $\alpha=-0.5$.

It must be stressed that the differences in SSFR obtained with various
attenuation laws are largely driven by the changes in SFR, rather than
$M_*$. The stellar masses of star-forming galaxies obtained with the
Calzetti dust attenuation law are on average only 0.06 dex lower than
the stellar masses obtained with the modified law. However, the masses
of a small number of individual galaxies, especially those with very
high SSFR, can differ up to 0.4 dex in either direction, i.e., the
dispersion of the two mass estimates increases with SSFR. For
completeness, we report that for passive galaxies the use of the
Calzetti dust law yields stellar masses that are 0.06 dex higher than
the ones obtained with our modified attenuation law, and this
difference is rather constant from one passive galaxy to another,
i.e., it has a small dispersion of only 0.02 dex.

We conclude that while star-forming galaxies may be governed by a
range of attenuation laws, they generally exhibit a steeper law than
the standard Calzetti curve ($\delta=-0.6$, on average), and include a
UV bump (1.1 times the MW value, on average). This average curve has
$R_V\approx 2.5$, compared with $R_V=4.05$ for the Calzetti curve, and
agrees well with the curve derived empirically by \citet{conroy10uv}:
a MW-like curve with $R_V=2.0$ instead of $R_V=3.1$ (reducing $R_V$
makes the curve steeper). Evidence for a steeper-than-Calzetti curve
has been found in other studies as well
\citep{cf00,buat11,wild11,hao11,salmon15}, though not as steep as the
one found here and in Conroy et al. More detailed investigation of
these results and their implications will be presented in a separate
paper.

\subsection{Correction for emission lines}

Finally, we describe how we account for the flux from emission lines,
which can have a significant effect on broadband fluxes and colors of
galaxies with high equivalent widths
\citep{papovich01,k03a,pacifici15}.  CIGALE calculates the
contribution of 124 lines, specified by metallicity (taken to equal
the stellar metallicity of SPS models) and ionization parameter. We
select the lowest ionization parameter available, $\log\, U=-3$, as it
produces the biggest improvements and is closest to direct
measurements \citep{dopita00,liang06}.

The formal quality of UV/optical fits is significantly improved by
correcting for emission-line flux (geometric mean of $\chi_{\rm
  red}^2$ drops from 1.2 to 0.7).  More importantly, accounting for
emission lines significantly improves the correlation of SED SSFRs
with other indicators. For galaxies with high SSFRs, not correcting
for emission lines produces offsets of up to 0.5 dex, as can be seen
by comparing Figure \ref{fig:ssfr_nolines}, which does not correct for
emission lines, with the lower row of Figure
\ref{fig:ssfr_comp}. Emission line correction makes very little
difference for stellar masses, which change by $<0.01$ dex on average.

While the addition of emission line to model fluxes results in overall
improvement, the correction is not perfect. Nebular metallicity is fixed to
stellar metallicity, which is not realistic. Also, the ionization
parameter of many galaxies is lower than the minimum ionization
parameter available in CIGALE. We see some redshift-dependent
trends in SFR, as lines straddle across the bandpasses.  We account
for this systematic offset ($\sim$0.1 dex on average) by deriving an
average correction, in 0.01 wide bins of redshift, with respect to
SFRs from B04 (using galaxies classified as star-forming; Section
\ref{ssec:comp_b04}), and applying it to all SFRs from the SED
fitting. We use B04 SFRs for this correction instead of mid-IR SFRs,
because the latter are available for only 63\% of star-forming
galaxies. Nevertheless, the correction would have been essentially
identical if mid-IR SFRs were used.

\subsection{Derived parameters}

In summary, for  each 0.01-wide redshift bin, from 0.01 to 0.30, we
calculate a grid of 342,720 models, a factor of three increase in
the number of models and a factor of 5 increase in redshift resolution
with respect to what was used in S07.

From the SED fitting we report the logarithm of the current stellar
mass ($M_*$), the logarithm of the SFR averaged over the last 100 Myr
(the timescale for UV emission), and dust attenuations in FUV, $B$ and
$V$ rest-frame bands. For each of these parameters CIGALE builds a
probability distribution function (PDF). The nominal value of the
reported parameter is the average of the PDF. We have performed mock
fitting, in which true parameters are known, and find that the average
of the PDF better retrieves the parameters than the median, and much
better than the parameter corresponding to the best-fitting model. The
best-fitting parameters are volatile and suffer from grid discreteness
(e.g., \citealt{taylor11}). The formal error of the parameter is taken
as the second moment (standard deviation) of the PDF. By default
CIGALE applies a minimum error in the case the error from the PDF is
lower than it, but we have disabled this adjustment.  Instead, we
caution the reader that the robustness of the error estimated from the
PDF will depend on the goodness of the fit, i.e., it will be most
accurate when $\chi^2_r\sim\!1$.

\begin{figure}
\epsscale{1.2} \plotone{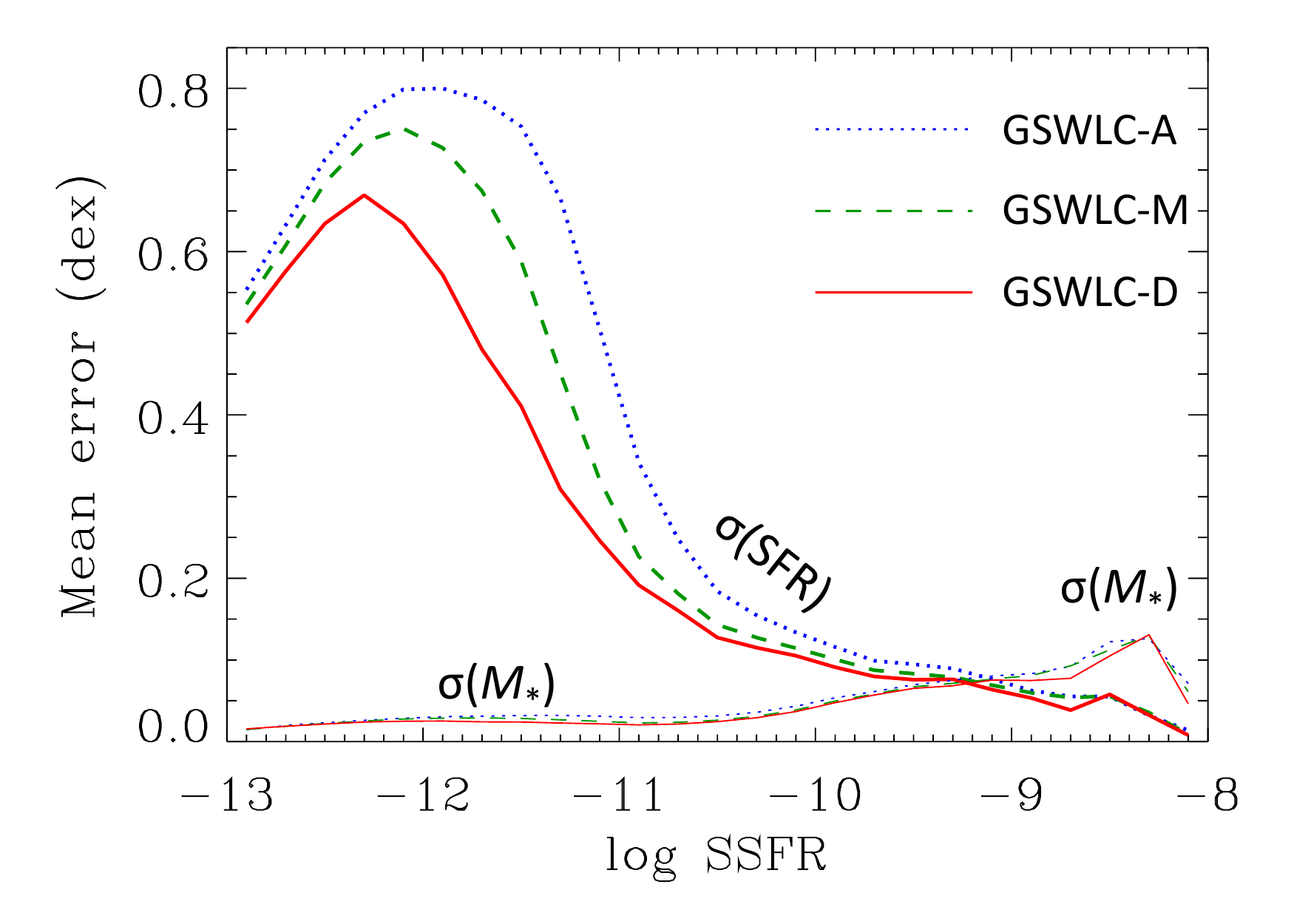}
\caption{Mean random error of SED fitting SFR (thick lines) and
  stellar mass (thin lines) as a function of SSFR and UV survey
  depth. The second moment of the parameter probability distribution
  function is adopted as the error of these
  parameters \label{fig:error}}
\end{figure}

\begin{deluxetable*}{l l l l} 
 \tablecaption{Contents of \galex--SDSS--\wise\ Legacy Catalog (GSWLC). \label{table:char}}
\tablewidth{0pt}
\tablenum{1}
 \tablehead{
   \colhead{Column number}      &
   \colhead{Column name}      &
   \colhead{Units} &
   \colhead{Description} \\
   }
\startdata
1 & ObjID &  & SDSS photometric identification number \\
2 & GLXID & & \galex\  photometric identification number \\
3 & plate & & SDSS spectroscopic plate number \\
4 & MJD & & SDSS spectroscopic plate date \\
5 & fiber ID & & SDSS spectroscopic fiber identification number \\
6 & RA & deg & Right Ascension from SDSS \\
7 & Decl.\ & deg & Declination from SDSS \\
8 & $z$ & & Redshift from SDSS \\
9 & $\chi^2_r$ & & Reduced goodnes-of-fit value for the SED fitting \\
10 & $\log M_*$ & $M_{\odot}$ & Stellar mass \\
11 & $\sigma(\log M_*$) & $M_{\odot}$ & Error of the stellar mass \\
12 & log SFR$_{\rm SED}$ & $\msperyr$ & UV/optical (SED) star formation rate \\
13 & $\sigma(\log {\rm SFR_{SED}} $) & $\msperyr$ & Error of the SFR \\
14 & $A_{\rm FUV}$ & mag & Dust attenuation in rest-frame FUV \\
15 & $\sigma(A_{\rm FUV})$ & mag & Error of dust attenuation in FUV \\
16 & $A_B$ & mag & Dust attenuation in rest-frame $B$ \\
17 & $\sigma(A_B)$ & mag & Error of dust attenuation in $B$ \\
18 & $A_V$ & mag & Dust attenuation in rest-frame $V$ \\
19 & $\sigma(A_F)$ & mag & Error of dust attenuation in $V$ \\
20 & flag\_sed & & SED fitting flag (0 = OK, 1 = broad-line spectrum,
2 = $\chi^2_r>30$, 5 = missing SDSS photometry) \\
21 & UV survey & & 1 = GSWLC-A, 2 = GSWLC-M, 3 = GSWLC-D \\
22 &  log SFR$_{\rm mid-IR, AW}$ &  $\msperyr$ & Mid-IR star formation
rate from \wise\ (AllWISE catalog)\\
23 & flag\_wise & & Mid-IR SFR (AllWISE) flag (0 = OK, 1 = no mid-IR
SFR (low SSFR), 5 = no 22 $\mic$ detection) \\
24 &  log SFR$_{\rm mid-IR, uW}$ &  $\msperyr$ & Mid-IR star formation
rate from \wise\ (unWISE catalog)\\
25 & flag\_unwise & & Mid-IR SFR (unWISE) flag (0 = OK, 1 = no mid-IR
SFR (low SSFR), 5 = no 22 $\mic$ detection) \\
26 & flag\_mgs & & 0 = not in SDSS Main Galaxy Sample (MGS), 1 = in
MGS \\
\enddata
\tablenotetext{}{Columns 10-19 originate from the SED fitting. If
  there are multiple reasons for setting the flag, the flag value wil
  be the sum of individual flag values. When the SED (or un/wise) flag
  is set, the SED fitting parameters (or mid-IR SFR) are not
  given. Mid-SFRs based on unWISE are recommended over the AllWISE
  ones for $z<0.06$ samples, large ($r>10\arcsec$) galaxies, or studies
  that explore dependence of SFR on galaxy size or shape. SFRs and
  stellar masses are based on Chabrier IMF. Missing values are listed
  as -99.}
\end{deluxetable*}

\section{GSWLC}\label{sec:catalog}

In Table 1, we describe the contents of the catalog which includes
various IDs, coordinates, the SED fitting parameters (SFR, $M_*$ and
dust attenuations in several bands), mid-IR SFRs and flags. Flags
describe cases when the SED parameters or the mid-IR SFRs are not
listed, as we discuss further below.

To review, the criteria for inclusion in the catalog are for an SDSS
object to be covered by \galex\ observations of a certain depth,
regardless of whether it was detected in the UV. Furthermore, galaxies
need to lie in $0.01<z<0.30$ redshift range and be above $r_{\rm
  petro}=18.0$ magnitude threshold.

Approximately 1\% of SDSS targets have spectra classified as
``quasars'' ({\tt CLASS = 1}), based on the Balmer emission lines being
broad. These galaxies are therefore likely to be Type 1 AGN, which
includes Seyfert 1s, and, less frequently, true quasars. In Type 1 AGN
the accretion disk is not entirely obscured and can contribute to the
UV/optical continuum of the host, which may significantly bias the
derived SED parameters, especially the SFR. This is confirmed by the
finding that the geometric mean of the reduced $\chi^2$ values of the
best-fitting model of broad-emission line galaxies is 4 times higher
than of other galaxies. We flag such objects in the final catalog and
do not report their SED-fitting parameters. Type 1 AGN are not
obscured by dusty torus, which is why they mostly contribute to the
UV/optical SED. We thus leave their mid-IR SFRs in the catalog, but
they should still be used with caution in case there is some AGN
emission in the mid IR.

We produce three separate catalogs, GSWLC-A, M, and D, according to
the UV depth, as described in Section \ref{ssec:sample}.  In addition,
we produce a master catalog that combines the three catalogs, keeping
for each galaxy the data from the deepest catalog. This catalog is
designated GSWLC-X, and contains 658,911 objects. Readers are
cautioned that non-uniform depth may lead to systematics in SFRs,
especially for galaxies with log SSFR$_{\rm SED}<-11$ where the UV
detection rate and the resulting quality differ from survey to survey
(Figure \ref{fig:error}). On the other hand, the stellar masses will
not be subject to any such biases.  All catalogs have the same
format as described in Table 1.  

\begin{figure*}
\epsscale{1.15} \plotone{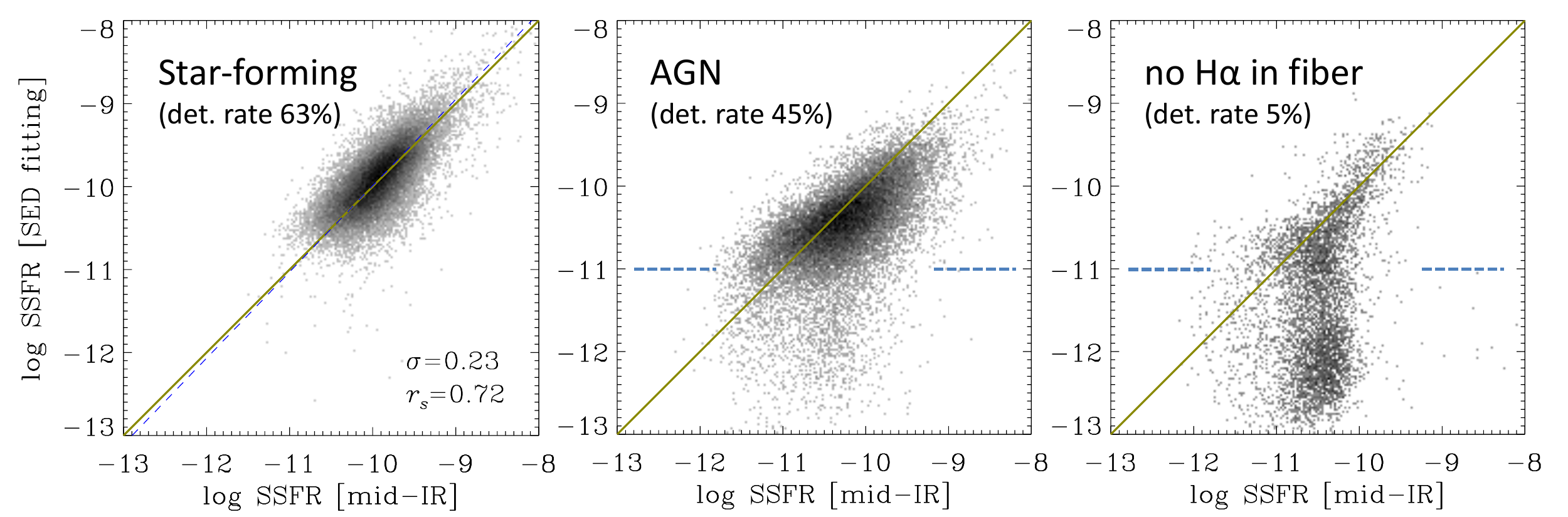}
\caption{Comparison of UV/optical (SED fitting) and mid-IR
  SSFRs. Mid-IR SSFRs are derived from the 22 $\mic$ mid-IR photometry
  from \wise\ (AllWISE catalog). Galaxies are split into SF and AGN
  classes according to the position in the BPT emission line
  diagram. The AGN category includes all galaxies on the AGN branch
  (i.e., including the galaxies that B04 calls SF/AGN composites and
  LINERs). Galaxies with lines too weak to allow classification are
  presented separately in the right panel. The left panel plots SF
  galaxies and shows the best fitting line along with its scatter and
  correlation coefficient.  Galaxies below dashed lines in middle and
  right panels have greatly overestimated mid-IR SFRs because their
  dust is primarily heated by old stars, invalidating the fixed
  $L_{\rm IR}$--SFR conversion. Their mid-IR SFRs have been removed
  from GSWLC. Both SSFRs use the stellar mass from our SED
  fitting. Detection rate at 22 $\mic$ of a given class of galaxy is
  given in each panel. Data in this and subsequent figures is from
  GSWLC-M. \label{fig:comp_wise}}
\end{figure*}

The catalogs are currently publicly available at
\url{http://pages.iu.edu/~salims/gswlc}, which will document any
changes implemented in subsequent versions. The catalog letter
designation is followed by a version number. The analyses in this
paper are all based on Version 1 catalogs (e.g., GSWLC-M1). GSWLC
is also hosted at Mikulski Archive for Space Telescopes (MAST) and may
be included in SDSS SciServer and/or hosted by NED/IPAC in the
future.

In Figure \ref{fig:error} we show mean random errors of log SFR$_{\rm
  SED}$ and log $M_*$ as a function of SSFR, for each of the three
catalogs GSWLC-A, M, D. Errors in SFR depend very strongly on the
SSFR, and for more passive galaxies also on the UV depth. The decrease
of errors at lowest and highest SSFRs is an artifact of reaching the
boundary of the model grid.  For passive galaxies the formal SFR
errors range between 0.65 and 0.80 dex. However, the SFR error of
passive galaxies that have no current SF whatsoever (log
SSFR$-\infty$) is infinity (in log). SFR values for galaxies with log
SSFR$<-11.7$ ($<-11.5$ for GSWLC-A; $<-12.0$ for GSWLC-D), should be
considered upper limits. For actively star-forming galaxies the SFR
errors are typically below 0.1 dex, a 50\% improvement over
S07. Errors on stellar mass are typically much lower than the SFR
errors, with the opposite dependence on SSFR, and little difference
between the UV surveys. They range from 0.03 dex for the passive
galaxies, to 0.10 dex for the most active ones.

The geometric mean of the reduced $\chi^2$ values for the GSWLC-M
catalog is 0.7. However, the tail extends to $\chi^2_r $ values in
excess of 100. Instead of introducing an arbitrary cut to exclude
poorly fit objects, we test the robustness of the derived SFRs and
stellar masses as a function of the reduced $\chi^2$. We find that no
systematic differences in excess of 0.1 dex (with respect to mid-IR
SFRs and stellar masses from B04) are present when $\chi^2_r \leq
30$. For those galaxies we retain the physical parameters from the SED
fitting, and annul them if $\chi^2_r>30$ (0.7\%). Nevertheless, the readers
are advised to treat galaxies with $\chi^2_r\geq 5$ with caution,
especially for individual galaxies.

\section{Comparison between SSFRs from the SED fitting, mid-IR,
 and \ha\ emission} \label{sec:comp}

In this section we discuss comparisons between various SFRs derived in
this work: specifically, the UV/optical SSFRs from the SED fitting on
the one side, and mid-IR and \ha\ SSFRs on the other. We remind the
reader that the SED (S)SFRs are averaged over the preceding 100 Myr,
the UV emission timescale.

The paper breaks away from the usual practice of comparing absolute
SFRs, and instead performs the comparison in terms of the specific
SFRs. The standard SFR comparison has its merits and may be more
intuitive, however, when the focus is on galaxies with similar stellar
populations, subject to similar systematics, the SSFR comparison has
its advantages: (1) SDSS probes a large dynamic range of distances,
and therefore of luminosities, which means that SFRs (and stellar
masses) will be correlated even if SFR/$L$ ($M_*/L$) are not. This
``trivial'' dependence is eliminated in SSFR. For example, the
correlation coefficient between SED and mid-IR SFRs (of galaxies
classified as star-forming), is $r_s=0.86$, while it is $r_s=0.72$ for
comparison involving SSFRs.  Consequently, the SFR comparison is
seemingly tighter. (2) The comparison of SSFRs is more informative
because it contrasts galaxies which are physically similar. For
example, a galaxy with SFR$=10 \msperyr$ can be a massive galaxy on
the star-forming sequence, or a lower-mass galaxy experiencing a
starburst. However, their SSFRs will be different. Or, a galaxy with
SFR$=0.01 \msperyr$ can be a blue star-forming dwarf, or a massive red
galaxy, but their SSFRs will differ by two orders of
magnitude. Finally, (3) the comparison in terms of SSFR allows for a
meaningful comparison between total and fiber SF, as long as the fiber
SFR is normalized by the stellar mass within the fiber.

\begin{figure*}
\epsscale{1.15} \plotone{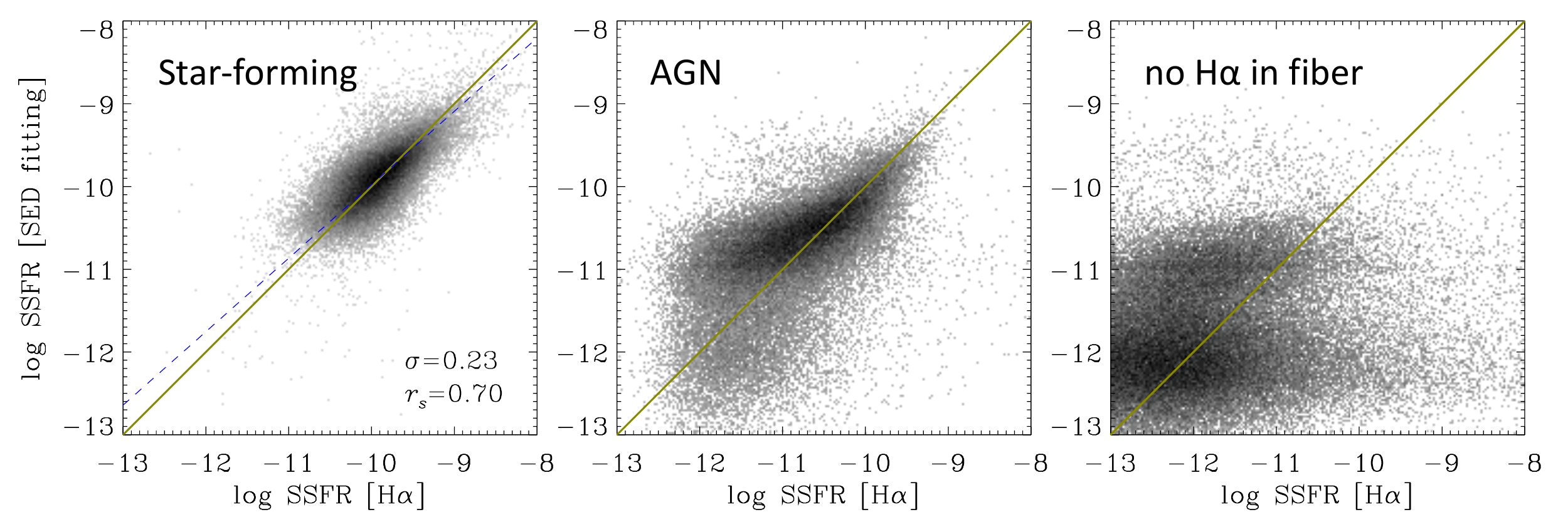}
\caption{Comparison of UV/optical (SED fitting) specific SFRs against
  SSFRs from dust-corrected \ha emission. SED fitting SSFRs are
  integrated (total), while \ha\ SSFRs are measured within SDSS fiber,
  covering on average 30\% of galaxy mass. Galaxies are split using
  the BPT diagram as in Figure \ref{fig:comp_wise}. The left panel
  plots SF galaxies and shows the best fitting line along with its
  scatter and correlation coefficient. \ha\ SSFR is normalized by
  stellar mass within the fiber. \label{fig:comp_ha}}
\end{figure*}

In Figure \ref{fig:comp_wise} we present the comparison of UV/optical
SSFRs (i.e., SED fitting SSFRs) against mid-IR SSFRs from \wise\ 22
$\mic$ photometry from AllWISE. We show the comparisons using medium
UV depth catalog (GSWLC-M), which balances the extensiveness of
GSWLC-A with the depth of GSWLC-D.  All of the findings hold for the
more accurate GSWLC-D as well. We refrain from using GSWLC-X, because
of its non-uniform UV depth. The comparisons are presented for
galaxies split into three categories using the BPT diagram ({\tt
  iclass} in MPA/JHU catalog). The star-forming class ({\tt iclass} =
1) includes galaxies with well-measured BPT lines and lying below the
\citet{k03c} empirical demarkation between galaxies with and without
AGN contribution. As we have already seen in Section \ref{sec:sed}),
once we have adopted an adequate attenuation law the comparison is
quite linear (slope in log-log plot equals 1), with no significant
systematics. Mid-IR SSFRs from \wise\ are available for 63\% of
galaxies in this class. The scatter around the bisector linear fit
(linear in log) is 0.23 dex, to which the measurement errors of SED
SFR and mid-IR SFR contribute approximately equally (0.15 dex
each). Note that in all of the figures where we show the linear fit,
we use a robust (outlier-resistant) bisector least square fit, which
treats the variables symmetrically, i.e., no variable is considered
independent.

Had we used mid-IR SSFRs based on unWISE photometry, the comparison
would look similar, but would have 5\% greater scatter. We find a
similar level of degradation with respect to SSFRs based on \ha. We
have not positively identified the cause of the increased scatter of
SSFRs based on unWISE compared to AllWISE photometry, but list two
possible reasons.  unWISE uses $r$-band profiles to extract
photometry. Optical light in $r$ band is dominated by the emission
from older stellar populations, so it may not represent an optimal
prior for 22 $\mic$ observations, where the bulge emission is
suppressed, while the star-forming regions in the disk
dominate. Another possibility is related to the fact that unWISE is
based on {\tt cmodelmag} profiles, whereas all other SFRs are based on
{\tt modelmag} SDSS photometry.  Despite being somewhat noisier in
comparisons with other (S)SFRs overall, unWISE IR luminosities and
SFRs are less biased for large ($>10\arcsec$) galaxies (where the
AllWISE PSF magnitudes can underestimate the SFR by up to 0.1 dex), so
we retain them in GSWLC alongside SFRs from AllWISE catalog. Our
recommendation is to use unWISE SFRs in studies that explore
dependence on galaxy size or profile (e.g., Sersic index) and also in
studies that focus on galaxies with large angular diameter. Otherwise,
AllWISE SFRs are recommended, and we continue to focus on them in
subsequent discussion.

The middle panel shows the comparison for AGN-hosting galaxies. This
category includes all galaxies above the \citet{k03c} line, i.e., what
B04 call SF/AGN composites, AGN (Seyferts) and LINERs ({\tt iclass} =
3, 4, 5). AllWISE detects 45\% of such galaxies. First, we notice that
no correlation is present in the lower part of the plot, approximately
when log SSFR$_{\rm SED}<-11$. These are nearly quiescent galaxies of
LINER type (thus, potentially passive galaxies with non-AGN emission
lines \citealt{stasinska08}). For such galaxies the IR emission, if
detected, will be dominated by dust heating by relatively dusty old or
intermediate-age (e.g.\ AGB) stellar
populations\citep{bressan01,villaume15}, and will therefore not be
indicative of the current SF \citep{s09,cortese08}. When the IR
luminosity of such galaxies is converted into SFRs using simple,
fixed-coefficient formulae (Eq. \ref{eq:lir}), which assume dust is
heated by young stars, the SFR will be overestimated
\citep{buat96,kennicutt98,boquien16}.  For AGN galaxies with higher
SSFRs, the mid-IR SSFRs tend to be up to 0.6 dex higher than the SED
SSFRs, suggesting a non-negligible dust heating by AGN, that produces
excess emission in the mid-IR.  The excess appears to be greater for
increasing SSFRs, which would be expected if the gas both fuels the SF
and drives the AGN accretion (e.g., \citealt{kewley06}).

Classification of a galaxy as a star-former or AGN, as performed by
B04, requires minimum adjusted SNR of 3 in \ha\ and other lines
($\approx 7.4$ in raw \ha\ SNR). Galaxies that are too weak to allow
classification, what we call ``no \ha'' category ({\tt iclass} = --1),
are shown in the right panel. These galaxies are mostly quiescent, so
it is not surprising that the AllWISE detection rate is only 5\%. When
detected, the majority of galaxies in this class have low SED SSFRs,
as expected. However, mid-IR SSFRs can be too high by up to 2 dex,
because the IR emission from old stars is interpreted as current
SF. We conclude that mid-IR SFRs are not reliable for quiescent or
nearly quiescent galaxies when obtained through simple recipes that
have a fixed conversion factor between IR luminosity and SFR
(Eq. \ref{eq:lir}), and therefore in GSWLC we remove mid-IR SFRs for
galaxies for which log SSFR$_{\rm SED}<-11$ (dashed line in middle and
right panels of Figure \ref{fig:comp_wise}).

Interestingly, there are some galaxies in the ``no \ha'' class for
which the SED fitting SSFRs are quite high, overlapping with the
values of normal star-forming galaxies (log SSFR$>-10.5$). It may seem
contradictory that a galaxy with essentially no \ha\ emission could
have such high SSFR. There are two possible reasons for this: (1)
Difference in SF timescales. \ha\ is not present because a galaxy is
in a post-starburst phase -- UV and mid-IR emission will persist after
the O-stars that give rise to \ha\ emission have died off. In other
words, high SED and mid-IR SSFRs reflect recent, but not instantaneous
values. The post-starburst galaxies would form the tail of high SSFRs
in Figure \ref{fig:comp_wise} (right; log SSFR$>-10$). Their SED and
mid-IR SSFRs agree, with no mid-IR excess like the one seen for
AGNs. This suggests that post-starburst galaxies, which are usually
considered to be the results of mergers (e.g., \citealt{yang08}),
nonetheless do not have a significant AGN emission.  (2) \ha\ is not
present within the SDSS fiber, but the SF (and presumably the \ha\
emission) is present outside of it. \citet{s12} and \citet{fang12}
have studied such population in detail with high-resolution UV,
optical and \ha\ imaging, and have confirmed that these are typically
lenticular (S0) galaxies with no SF in the bulge, but with low-level
SF present in a ring outside of SDSS fiber. The SSFRs of such
star-forming S0s would be in the range $-12<$ log SSFR $<-10$, i.e.,
lower than post-starburst galaxies.

Figure \ref{fig:comp_ha} shows equivalent comparison of SED SSFRs with
respect to \ha\ SSFRs. \ha\ SSFRs are measured within the $3\arcsec$
SDSS fiber, which covers between 17\% and 50\% of the galaxy's stellar
mass. The comparison of star-forming galaxies (left panel) shows
similar scatter as that seen with respect to the mid-IR. The relation
is somewhat SSFR dependent. One expects the relation between total and
fiber SSFR to be linear (slope in log-log plot equal to 1) only if
there are no SSFR gradients. Somewhat increased nonlinearity for galaxies
with log SSFR$_{{\rm H}\alpha}\geq-9$ suggests either that the SF tends to be
centrally concentrated so that it produces higher SSFRs within the
fiber compared to the integrated SSFR, or, that the fiber mass from
the MPA/JHU catalog, which we use to normalize the \ha\ SFR, is
systematically underestimated for bursty galaxies (Section
\ref{ssec:comp_mass}).

For galaxies with AGN/LINER contribution (middle panel), there are no
systematic differences for hosts having high SF (log SSFR$>-10$). This
is in contrast to mid-IR SSFRs that showed an excess in this
regime. \ha\ SSFRs start to display systematic discrepancies at lower
SSFRs, in the sense that \ha\ SSFRs are up to 0.7 dex too low. We find
that many of such cases are galaxies with no optical signs of SF,
either photometrically (optical colors are red) or spectroscopically
(not much \ha), and yet the UV is clearly indicative of SF--often the
UV colors are noticeably blue and uniformly spread across the
disk. These galaxies may be in a declining phase of SF where the UV
emission is higher than the rapidly diminishing \ha.

For galaxies with "no \ha'' (right panel), no correlation is present,
as expected given the weakness of \ha\ that defines this category and
making \ha\ SSFRs essentially meaningless.  There is an overdensity of
galaxies below log SSFR$_{\rm SED}=-12$, where SED SFRs estimates also
become rather uncertain (Figure \ref{fig:error}) and are more likely
just the upper limits of galaxies with no current SF whatsoever, such
as the majority of early-type galaxies.

To conclude, for actively star-forming galaxies all three indicators
(SED fitting, mid-IR, and \ha) provide robust measurements of
(S)SFR. For mid-IR, the depth of \wise\ allows only 2/3 of SDSS
galaxies to be detected.  \ha\ (S)SFRs pertain only to the fiber.
Mid-IR luminosity breaks down as a SFR indicator for galaxies with log
SSFR$_{\rm SED}<-11$, where dust heating is dominated by old
populations. This corresponds to green-valley and quiescent
galaxies. Furthermore, the mid-IR SFR appears to be affected by AGN
emission, limiting its usefulness. SED SSFRs are measurements of
choice in low-SSFR regime, but even they become essentially upper
limits when log SSFR$_{\rm SED}<-11.7$ (for GSWLC-M; -12.0 for
GSWLC-D; -11.5 for GSWLC-A).

\section{Comparison of GSWLC with previous catalogs} \label{sec:other}

In this section we compare GSWLC stellar masses and SED fitting SFRs
(primarily through SSFRs) with those from previously published
catalogs. The focus will be on (S)SFR comparisons, which are more
sensitive than stellar mass measurements. We continue to present the
comparisons using GSWLC-M, but the results hold with GSWLC-A and
GSWLC-D.

\begin{figure}
\epsscale{1.2} \plotone{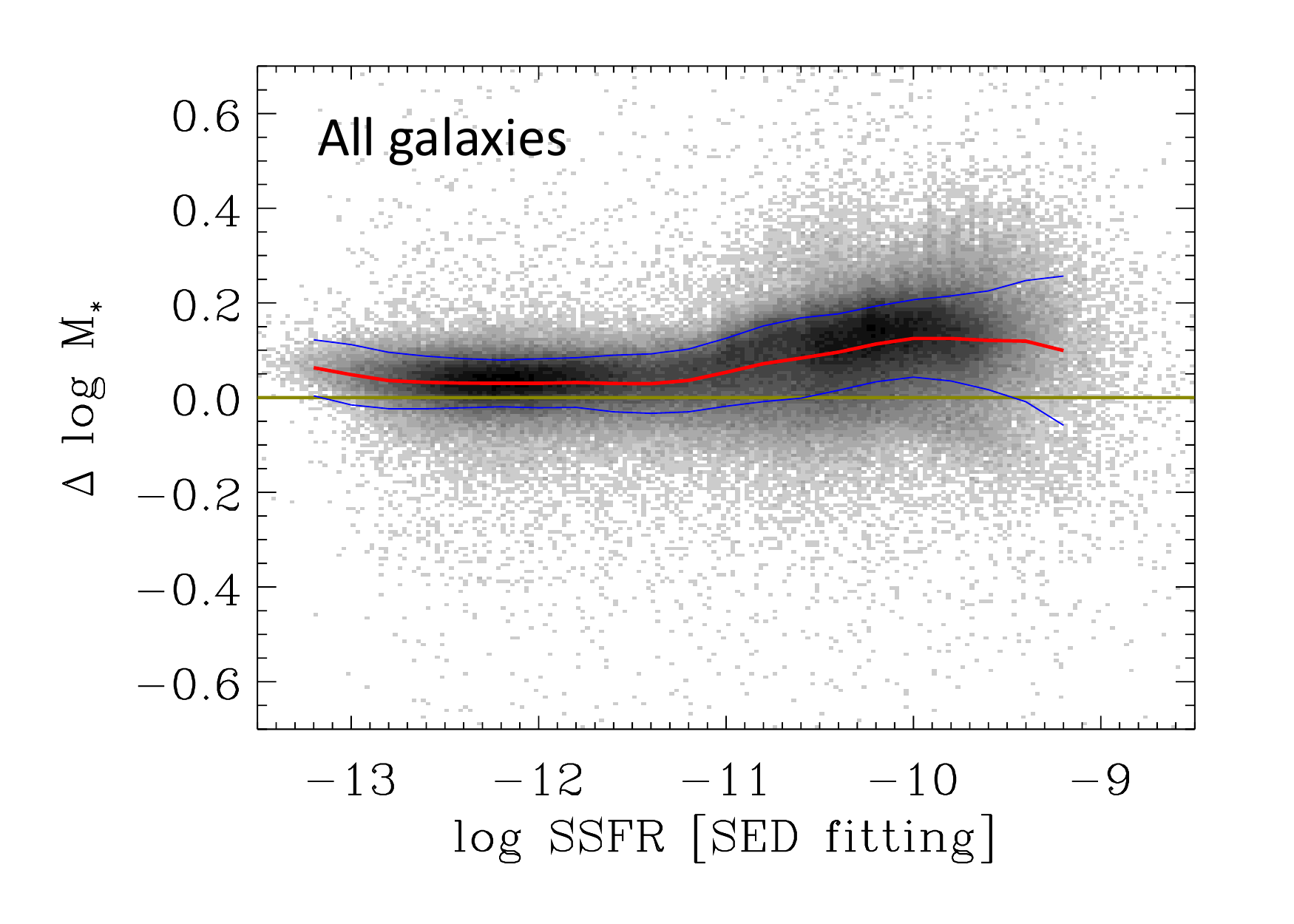}
\caption{Comparison of stellar masses from GSWLC-M and MPA/JHU (DR7)
  catalogs, presented as the difference in masses as a function of
  SSFR. Red curve are the averages of the difference in 0.2 dex wide
  bins, and the blue curves are the averages $\pm 1 \sigma$. GSWLC
  masses are on average higher than MPA/JHU catalog masses because of
  the differences in the assumed SF histories, but the difference is
  quite small (0.03--0.13 dex), and is SSFR dependent. The standard
  deviation of the mass difference is typically 0.07
  dex. \label{fig:comp_mass}}
\end{figure}

\begin{figure*}
\epsscale{1.15} \plotone{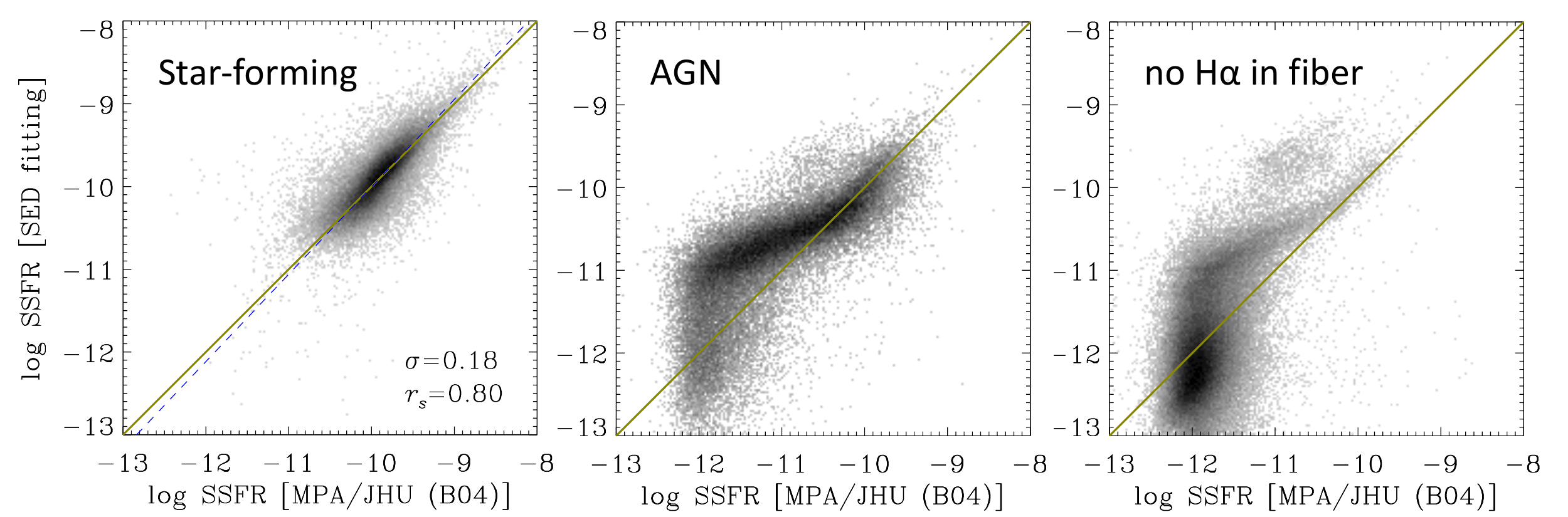}
\caption{Comparison of SSFRs from the SED fitting against
  aperture-corrected (total) SSFRs from the MPA/JHU DR7 catalog, derived
  based on a modified \citet{b04} method. Galaxies are classified using the
  BPT diagram, as in Figure \ref{fig:comp_wise}. SF class shows the
  best fitting line and its scatter and correlation
  coefficient. Agreement is excellent for SF class, but systematics
  are present for low SSFR AGN and ``no \ha'' galaxies for which
  MPA/JHU SFRs are based on indirect methods that are not sensitive to
  low levels of SF.\label{fig:comp_b04}}
\end{figure*}

\subsection{Comparison with MPA/JHU stellar masses} \label{ssec:comp_mass}

The MPA/JHU catalog is the most widely used source of stellar masses
for SDSS galaxies. Prior to DR7, the most recent version of this
catalog, the MPA/JHU catalog listed stellar masses derived in
\citet{k03a}, using a method that combines photometric information to
assess dust attenuation, and spectroscopic indices to constrain the SF
history and mass-to-light ratio. Spectroscopic indices were measured
within the fiber, so the method assumed no $M/L$ gradients. In the DR7
version of MPA/JHU catalog, the method from Kauffmann et al.\ was
replaced with the Bayesian SED fitting method, using only the optical
broad-band photometry from SDSS ({\tt modelMag}) and models described
in S07. Here we will only present comparisons with respect to DR7
version of the MPA/JHU catalog.

As described in Section \ref{sec:sed}, GSWLC uses different
specification of SF histories and dust extinction from those used in
S07 and the MPA/JHU catalog. Furthermore, unlike the MPA/JHU catalog
(but, like S07), GSWLC uses the constraints offered by UV photometry.
Figure \ref{fig:comp_mass} shows the difference between GSWLC and
MPA/JHU stellar masses as a function of SSFR. Again, we prefer the
presentation of the results in this, relative, distance-independent
way than the usual mass vs.\ mass comparison.  On average, GSWLC
stellar masses are somewhat higher than the MPA/JHU ones, with typical
difference being 0.03 dex for passive galaxies, and $\sim$0.13 dex for
the active ones. The scatter of the difference in mass, 0.07 dex, is
consistent with the formal estimates of the mass error (Figure
\ref{fig:error}). The inclusion of the UV photometry is responsible
for 0.04 dex increase for active galaxies. The remaining 0.09 dex
increase for active galaxies (and 0.03 dex increase for passive ones)
is mostly due to our use of two-component exponential SF histories.
We confirm that the difference would not have been present if we had
used the {\it delayed} exponential histories instead (Eq.\
\ref{eq:delayed}). The critical difference between the delayed
exponential and our implementation of the two-component exponential SF
history is that in the latter the old component starts in the early
universe, whereas the delayed exponential, having a single peak, will
be shifted towards later epochs. Recently, \citet{sorba15} have
reported that the masses of nearby high-SSFR galaxies obtained by
summing up the masses in individual ``pixels'' are up to 0.1 dex
higher than the masses from integrated light. They proposed that the
difference arose from ``outshining'' of the old, fainter populations
in galaxies with younger populations (high SSFR). It appears that our
use of the two-component exponential SF history, in which the old
component is set to have started in the early universe, may have
recovered this deficit.

More recently, stellar masses for SDSS galaxies were published by
\citet{mendel14} and \citet{chang15}. The differences of GSWLC masses
with respect to these masses follow the trends we have shown with
respect to masses from the MPA/JHU catalog: little difference for
passive galaxies, and up to 0.2 dex difference for galaxies with high
SSFR.

Comparison of our masses obtained with and without the UV photometry
reveals that in the case when there is another photometric source
within 2--3$\arcsec$, the masses obtained from joint UV and optical
photometry will be biased upward by 0.05 dex, presumably because of
the unaccounted blending in the UV. This affects only a few percent of
all sources.

\subsection{Comparison with MPA/JHU star formation rates} \label{ssec:comp_b04}

The MPA/JHU catalog also provides SFRs, and until recently it was the
only publicly available source of SFRs for SDSS galaxies. SFRs derived
in MPA/JHU follow the method of B04, with modifications introduced in
the most recent (DR7) version of the catalog.  We first describe both
the original method and its modifications, but the comparison will be
presented only for the DR7 version of MPA/JHU catalog.

MPA/JHU catalog (and the original B04) SFRs are often described as
\ha, or emission-line SFRs.  This is accurate only for the portion of
a galaxy contained within the fiber, and only for galaxies classified
as star-forming using the BPT diagram, for which the AGN contribution
to emission lines should be negligible. For galaxies with AGN
contribution or having weak lines (altogether 78\% of SDSS galaxies),
B04 derive SFR in the fiber based on a relation between the
emission-line SSFR and D4000 index, constructed from star-forming
galaxies.  Next, the SFR estimate within the fiber is
aperture-corrected to produce the total (integrated) SFR. B04 performs
this correction by first establishing the relationship between fiber
SSFR and the fiber broad-band colors of star-forming galaxies, and
then applying these relations in a Bayesian fashion to the light
outside of the fibers to arrive at the out-of-fiber SFR. The total SFR
is then obtained as the sum of fiber and out-of-fiber SFRs. The DR7
MPA/JHU catalog modifies the procedure for deriving out-of-fiber SFRs,
by instead performing the SED fitting to $ugriz$ photometry, using the
models and methods described in S07. Altogether, the SFR method of
MPA/JHU catalog (and original B04) is an emisison-line/D4000/SED
fitting hybrid. The temporal sensitivity of such SFRs will be between
the $\sim\!10$ Myr timescales traced by the emission lines and $\sim\!
1$ Gyr timescale for $u$-band light.

S07 presented detailed comparison of their SED SFRs and the original
SFRs from B04.  GSWLC contains many improvements over the S07
methodology, as does the MPA/JHU catalog with respect to B04. The
comparison of GSWLC-M and DR7 MPA/JHU total SSFRs is given in Figure
\ref{fig:comp_b04}. It uses the same division of galaxies into SF, AGN
and ``no \ha'' categories as employed earlier. In order to decouple
SFR and mass systematics, both SSFRs are obtained by normalizing by
the same mass (from GSWLC).  Note that for 3\% of galaxies in MPA/JHU
catalog the total SFRs are present even when masses are not. We find
that in those cases the reported total SFR is incorrect, and assumes
the value of {\em fiber} SFR. We exclude those values from the
comparison.

SSFRs for star-forming galaxies (left panel) agree very well, with a
scatter of $\sigma=0.18$, which is smaller than the scatter of SED
SSFRs with respect to mid-IR or \ha\ SSFRs (Figure \ref{fig:comp_wise}
and \ref{fig:comp_ha}). The comparison for AGN hosts reveals some
systematic discrepancies, especially for intermediate SSFRs, which can
in some cases reach $\sim$1 dex. We have already seen similar, but
smaller offsets in comparison with \ha\ {\em fiber} SSFRs, which we
attributed to optically (spectroscopically {\it and} photometrically)
inconspicuous SF. If we intentionally leave out the UV bands from our
SED fitting, the resulting (ill-constrained) SSFRs are drawn to lower
values (because the majority of models with red optical colors has low
SSFRs), in better agreement with B04 values, which derive the greater
part of their SFR (the out-of-fiber portion) from similar optical-only
SED fitting.

Similar trends are present in galaxies with little or no \ha\ emission
in the fiber (right panel), except that the majority of galaxies have
low SSFRs (log SSFR$_{\rm SED}<-11$). In this class of galaxies there
exists a peculiar feature: a cloud of galaxies lying $\sim$1 dex above
the 1:1 line, having high SED SSFRs (log SSFR$_{\rm SED}>-10$), but
much lower MPA/JHU SSFRs. We confirm that these galaxies have robust
SED fits and are UV detected. In Section \ref{sec:comp} we mentioned
that ``no \ha'' galaxies with high SSFRs are probably post-starburst
galaxies. Such galaxies are traditionally identified as E+A galaxies,
based on their Balmer absorption features \citep{dressler83}. Now we
inspect the spectra of $\sim$200 of these outliers and confirm the
presence of Balmer absorption lines. Visually ($gri$ composites),
these galaxies appear like red early-type galaxies, often with white
centers, suggesting a central (post) starburst. We match our sample to
\citet{goto07} catalog of E+A galaxies (online version updated with
SDSS DR7 data) and confirm that the majority of their E+As are found
in this region of the plot. Why do the two methods yield discrepant
SFRs for these galaxies? We find that the SED fitting (S)SFRs of E+A
galaxies are very sensitive to the assumed dust attenuation law, more
so than the normal star-forming galaxies. The (S)SFRs that we obtain
for E+As assuming the modified attenuation law are an order of
magnitude higher than what would be derived using the nominal Calzetti
law. The reduced $\chi^2$s are 5 times lower when the modified
attenuation law is used. From this we conclude that the high recent
(S)SFRs of E+As that we derive in GSWLC are more likely to represent
true levels of SF averaged over the last 100 Myr.

\subsubsection{On the systematics of MPA/JHU SFRs reported by SAMI
  Galaxy Survey}

Recent campaigns to obtain resolved spectra with
integral-field spectroscopy have reported systematic differences
with respect to SFRs from the MPA/JHU catalog. In particular,
\citet{richards16}, using preliminary data from the SAMI Galaxy Survey
\citep{sami}, find a non-linear, i.e. SFR-dependent, relation between
their and MPA/JHU SFRs (which they refer to as B04 SFRs), in the sense
that galaxies with high SAMI SFRs have underestimated MPA/JHU
SFRs. The discrepancy is already $\sim\!0.3$ dex at log SFR$_{\rm
  SAMI}\approx 0.6$, the highest SFRs in their sample (we convert
all SFRs from Richards et al.\ to Chabrier IMF). Richards et al.\
attribute the discrepancy to possible biases in B04 methodology for
deriving aperture corrections.

In our analysis so far we have shown that GSWLC {\it specific} SFRs of
star-forming galaxies have no significant systematics with respect to
either mid-IR SSFRs or SSFRs from the MPA/JHU catalog. This implies,
and we confirm it to be true, that MPA/JHU and mid-IR SSFRs agree
between themselves as well. While this seems to imply that MPA/JHU
measurements are not biased, it is necessary to verify if such results
hold for SFRs, and not just the SSFRs.

\begin{figure}
\epsscale{1.05} \plotone{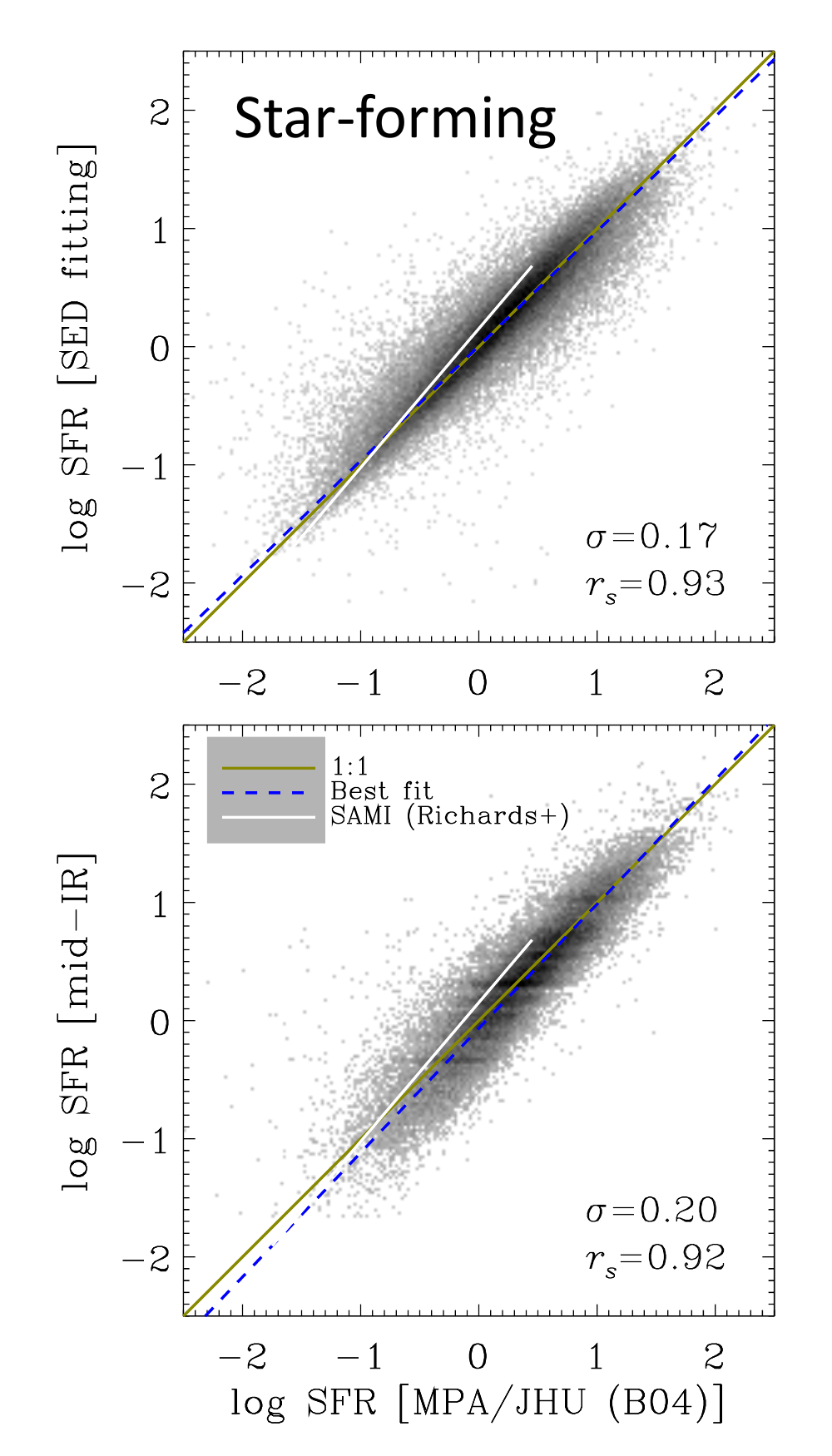}
\caption{Comparison of SFRs from the SED fitting (upper panel) and
  from \wise\ mid-IR (lower panel), with respect to SFRs from the
  MPA/JHU catalog \citep{b04}. Shown are the galaxies from GSWLC-M
  classified as star-forming on the BPT diagram. There are no
  systematic differences or significant non-linearities. This is in
  contrast with the relation derived from integral-field spectroscopy
  \ha\ measurements from SAMI Galaxy Survey \citep{richards16}, shown
  as the white solid line (SAMI vs. MPA/JHU). In this figure we lower
  the redshift limit to $z_{\rm min}=0.01$ to increase the
  contribution of low-SFR galaxies. Slight discretization in mid-IR
  SFRs arises from sampling of \citet{ce01} IR
  templates. \label{fig:sfr_b04_sami}}
\end{figure}

\begin{figure*}
\epsscale{1.15} \plotone{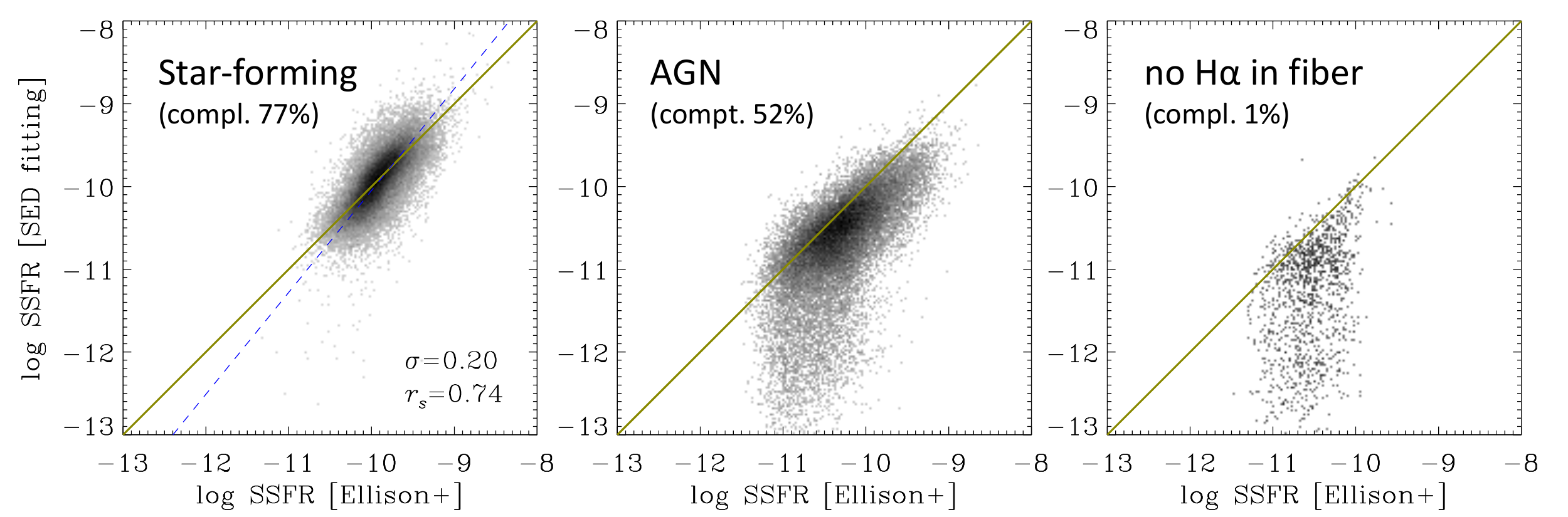}
\caption{Comparison of SSFRs from the SED fitting (from GSWLC-M) with
  SSFRs from the IR luminosity catalog of \citet{ellison16}, derived using
  artificial neural network (ANN) method trained on {\it Herschel+WISE}
  data.  Galaxies are classified using the BPT diagram as in Figure
  \ref{fig:comp_wise}. SF class shows the best fitting line and its
  scatter and correlation coefficient. Both SSFRs are normalized by
  the total stellar mass from GSWLC SED fitting. Comparison is similar
  to one involving mid-IR SSFRs (Figure \ref{fig:comp_wise}), with
  some non-linearity for star-forming galaxies (left
  panel). Completeness of \citet{ellison16} catalog for a given class
  is indicated in each panel \label{fig:comp_e16}}
\end{figure*}

\begin{figure}
\epsscale{0.9} \plotone{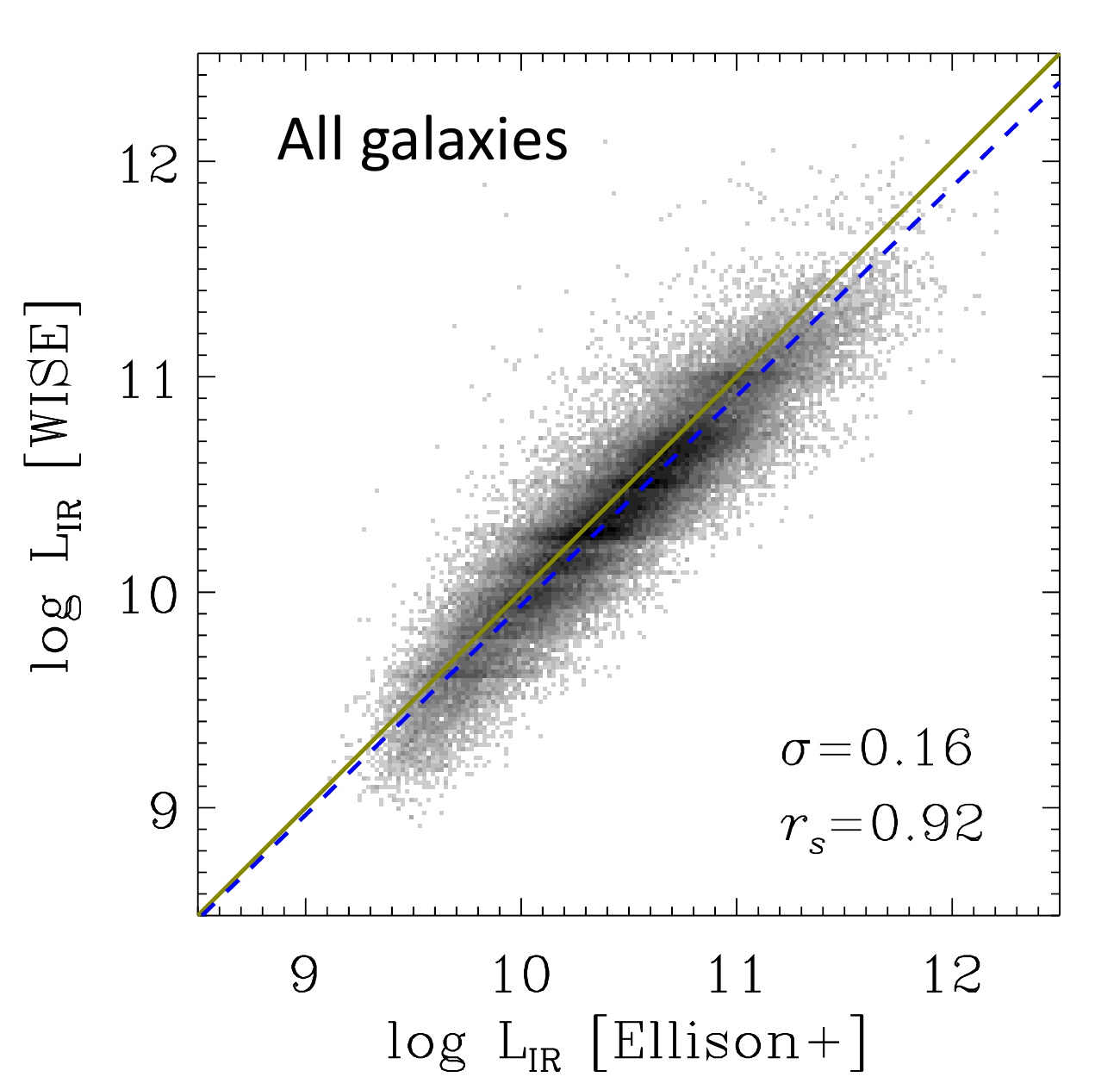}
\caption{Comparison of total IR luminosities derived in GSWLC using
  \wise against the total IR luminosities derived with ANN in
  \citet{ellison16}. \wise measurement is direct, but is based on
  extrapolating a single flux point at rest-frame $\sim\!20\mic$. ANN,
  on the other hand, uses various optical photometric and
  spectroscopic measurements to predict the IR luminosity based on a
  training set consisting of IR luminosities from {\it Herschel+WISE}
  (three sub-mm flux points from SPIRE and 22 $\mic$ point from
  \wise). The comparison is good notwithstanding some systematics for
  the most luminous galaxies. Luminosities are given in solar
  units. \label{fig:comp_lir_e16}}
\end{figure}

Figure \ref{fig:sfr_b04_sami} shows SED fitting SFRs against MPA/JHU
SFRs in the upper panel, and mid-IR SFRs against MPA/JHU SFRs in the
lower panel. In both comparisons, the linear fits (blue dashed lines)
follow closely the 1:1 relation. White solid lines show the relation
between SAMI and B04 SFRs (adjusted to Chabrier IMF), over the range
of SFRs covered in \citet{richards16}.\footnote{The highest SFRs in
  \citet{richards16} are an order of magnitude lower than the highest
  SFRs in SDSS, probably because their sample is drawn from much
  smaller volume compared to that of SDSS (smaller area, plus
  $z<0.06$.)} The discrepancies suggested byRichards et al.\ relation
are clearly excluded with either the SED fitting or the mid-IR SFR.
Results are unchanged when the redshift range of galaxies is
restricted to match the redshift range of SAMI Galaxy
Survey. Apparently, more work is needed to understand the source of
differences. For now, we conclude that the total MPA/JHU SFRs of
actively star-forming galaxies do not appear to be biased when
compared with two independent measures of integrated SFR.

\subsection{Comparison with \citet{ellison16} star formation rates}

All-sky far-IR observations that would enable direct measurement of
the total IR luminosity across SDSS are only available from relatively
shallow {\it IRAS} and {\it AKARI} surveys, which preferentially
more luminous galaxies \citep{ellison16}. In order to
produce IR luminosity estimates for a significant portion of SDSS
spectroscopic sample, \citet{ellison16} apply an artificial neural network (ANN)
technique, using IR luminosities from \citet{rosario16} as the
training set (these IR luminosities were discussed in Section
\ref{sec:wise}). The idea behind the ANN is to establish intrinsic
correlations between observable quantities of the training set and the
target quantity (in this case the IR luminosity) and then apply these
correlations to estimate (``predict'') the target quantity in the full
dataset. The robustness of ANN estimates should be tested by comparing
them with the independent measurements of the target quantity for an
unbiased subsample drawn from the full dataset. In the case of
\citet{ellison16}, the parameters employed to estimate the IR
luminosities using ANN include, in addition to the redshift, the
photometric (magnitudes, colors) and spectroscopic measurements
(emission lines strengths and D4000 break), as well as the stellar
mass.  The connection between the fiber and total quantities is
established by also including the mass in the fiber and the $r$-band
fiber covering fraction.  The input parameters come from MPA/JHU DR7
catalog.  The requirement to have available all of the parameters in
the target dataset limits the application to 45\% of SDSS galaxies,
typically the ones with stronger emission lines. IR luminosities of
the training set are recovered with the typical accuracy of $\sigma
\sim 0.1$ dex, with no major systematics \citep{ellison16}.

\begin{figure*}
\epsscale{1.15} \plotone{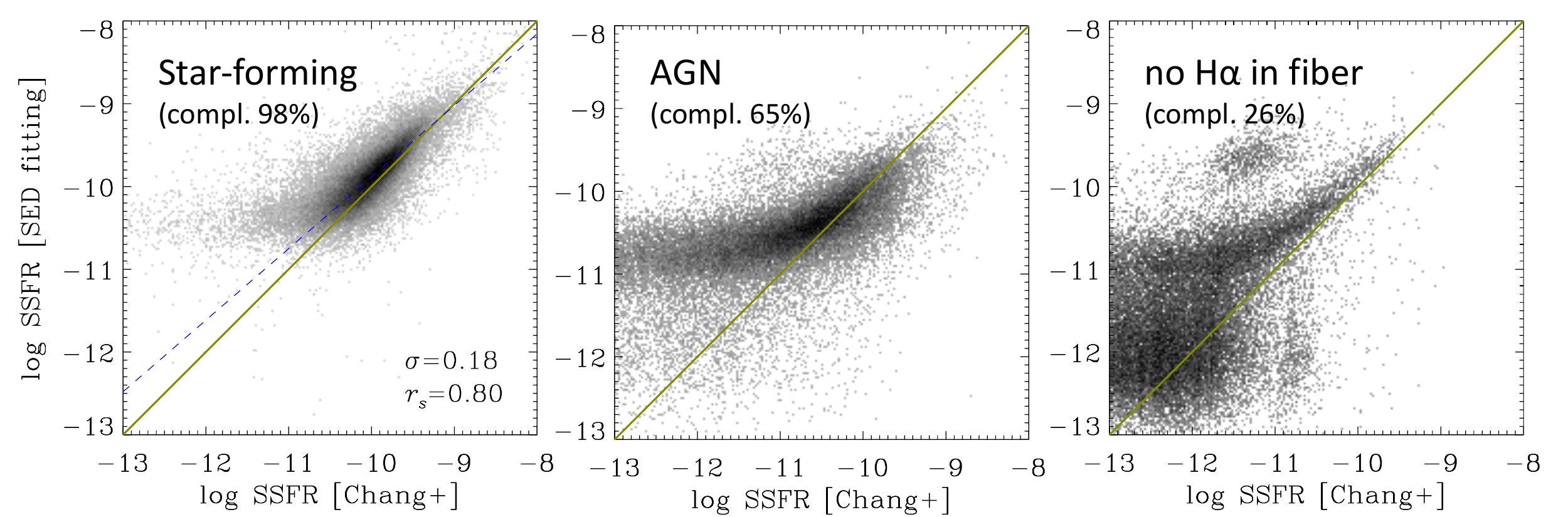}
\caption{Comparison of SSFRs from the SED fitting against the SSFRs
  from \citet{chang15}, derived from optical+mid-IR SED fitting using
  MAGPHYS \citep{dacunha08}.  Galaxies are classified using the BPT diagram
  as in Figure \ref{fig:comp_wise}. SF class shows the best fitting
  line and its scatter and correlation coefficient. Both SSFRs are
  normalized by the total stellar mass from GSWLC SED fitting. Low
  SSFRs tend to be underestimated in \citet{chang15} catalog, but
  otherwise the correlation is good with a small scatter. Completeness
  of \citet{chang15} catalog for a given class is indicated in each
  panel. \label{fig:comp_c15}}
\end{figure*}

\begin{figure}
\epsscale{0.9} \plotone{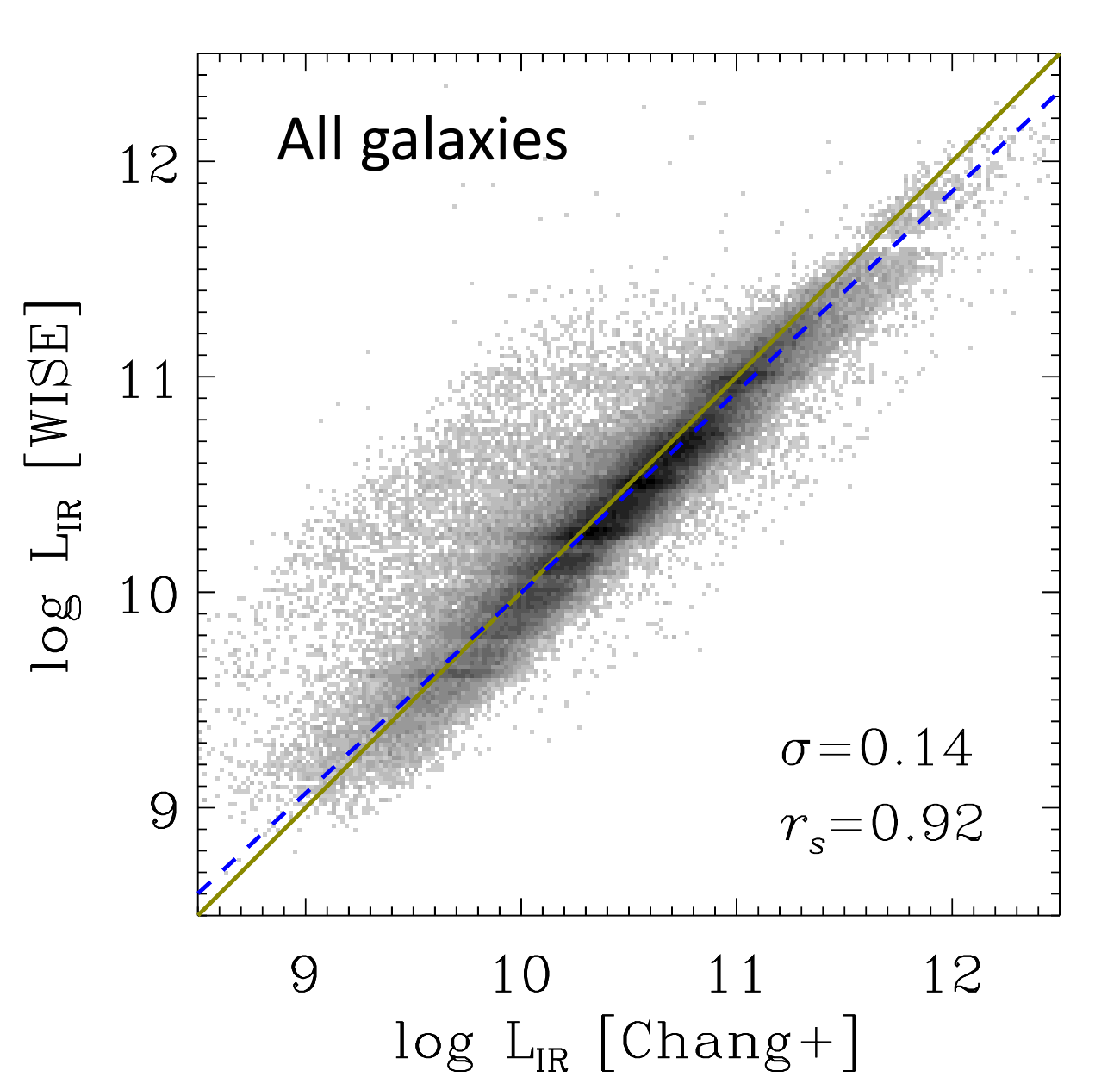}
\caption{Comparison of total IR luminosities derived in GSWLC from
  \wise\ to total IR luminosities derived from optical/mid-IR SED
  fitting from \citet{chang15}. The comparison is generally quite
  good. A tail to the left is due to a small number of passive
  galaxies for which Chang et al.\ luminosities are too
  high. \label{fig:comp_lir_c15}}
\end{figure}

In Figure \ref{fig:comp_e16} we show the comparison of GSWLC SSFRs
from the SED fitting against SSFRs converted from \citet{ellison16} IR
luminosities using Eq.\ \ref{eq:lir}. The comparison is limited to
galaxies with $\sigma_{\rm ANN}<0.1$, a cut recommended in
\citet{ellison16} to remove the galaxies whose estimated IR
luminosities may be uncertain. $\sigma_{\rm ANN}$ essentially measures
the degree to which a target galaxy is represented in the training
set. The larger the value, the less likely the target galaxy is
represented in the training set and its IR luminosity may therefore be
uncertain.  The application of the cut on $\sigma_{\rm ANN}$ decreases
the number of galaxies with ANN IR luminosities from 330,000 to
250,000, or 1/3 of SDSS.  The comparison of SSFRs of star-forming
galaxies (left panel) shows good general agreement, with mild
non-linearity (slope of 1.23), which can be traced to the differences
between \wise\ and {\it Herschel-WISE} specific IR luminosities
(Section \ref{sec:wise}). If the galaxies with $\sigma_{\rm ANN}>0.1$
were included in the comparison, the scatter at high SSFRs would
increase, presumably because such galaxies are rare in the training
set. ANN IR luminosities are available for 77\% of galaxies in this
class (after the application of $\sigma_{\rm ANN}$ cut), which is
higher than the \wise\ detection rate of 63\%.

For galaxies classified as AGN (the middle panel), the correlation is
present when log SSFR$_{\rm SED}>-11$, but with ANN SSFRs tending to
be higher, especially for galaxies with high SSFR. Below log
SSFR$_{\rm SED}=-11$ there is no correlation. Both of these behaviors
mimic the comparison of SED fitting SSFRs with mid-IR SSFRs (Section
\ref{sec:comp} and Figure \ref{fig:comp_wise}, middle panel). Excess
SSFR for AGNs with higher SSFRs, which in the case of mid-IR SSFRs we
attributed to AGN dust heating, is somewhat surprising, because it
would be expected that the AGN contribution, which drops above 40
$\mic$ \citep{mullaney11}, will be significantly diminished in the
total IR luminosity. However, we remind the reader that ANN
luminosities are trained on IR luminosities derived from a combination
of 22 $\mic$ flux (which is subject to AGN contamination) and three
sub-mm flux points that lie well beyond the IR SED peak (and are
therefore not as sensitive to the current SF, and more to the cold
dust mass).  More detailed analysis on the contribution of Type 2 AGN
to IR SED lies outside of the scope of this paper.

After the application of $\sigma_{\rm ANN}$ cut, the IR luminosities
from \citet{ellison16} are available for only 1\% of galaxies having
weak or no \ha\ (right panel). For a handful of such objects with high
SSFRs (including confirmed E+As) the match between the SED fitting and
ANN SSFRs is good, but for the great majority, the ANN IR
luminosities, when interpreted as SFRs, tend to be too high because of
the dust heating from older stars, as already discussed in Section
\ref{sec:comp}.

Considering the very different methods of estimating the IR
luminosities, it is interesting to see how the ANN IR luminosities
from \citet{ellison16} compare to the ones we obtain from \wise. This
is shown in Figure \ref{fig:comp_lir_e16}, for all galaxies for which
the two measurements are available (and with $\sigma_{\rm ANN}<0.1$
cut applied). The comparison, extending over three orders of
magnitude, is fairly good, both in terms of offset and the scatter
($\sigma=0.17$). Some systematic differences are present, especially
at $\log L_{\rm IR}>11$, which is not surprising considering that the
training set peters out at those luminosities. Overall, the IR
luminosities from \citet{ellison16} represent remarkably good
estimates considering that they were determined from optical
properties alone, but their use as SFRs is subject to the same caveats
(AGN dust heating and breakdown for passive galaxies) as the SFR
obtained from \wise\ mid-IR data.

\subsection{Comparison with \citet{chang15} star formation rates}

In Section \ref{sec:wise} we mention that new SED fitting codes allow
the modeling of the SED to extend into the IR, by including the dust
emission. This approach is used in MAGPHYS \citep{dacunha08}, which
models the IR SED as a sum of various SED components, the relative
contribution of which is mildly related to galaxy's SSFR.  CIGALE
allows the IR SED to be modeled according to one of the four published
template sets, without constraints on the shape. In both cases the
dust luminosity (i.e., the total IR luminosity) is normalized to match
the stellar emission absorbed in the UV/optical/near-IR. Such IR
luminosity will therefore include dust heating from stars of all ages.

Stellar plus dust emission modeling is applied in
\citet{chang15}, who use MAGPHYS to perform the SED fitting
simultaneously on optical ($ugriz$) photometry from SDSS and mid-IR
photometry from \wise\ (3.4, 4.6, 12 and 22 $\mic$). Of these 9 bands,
all but the longest two \wise\ bands will be dominated by the stellar
emission. Emission at 12 and 22 $\mic$, if detected, will help
constrain the dust luminosity and therefore the SFR (if the dust is
primarily heated by young stars).

We follow the analysis established in previous sections, and in Figure
\ref{fig:comp_c15} present comparisons between our SED fitting SSFRs
and SSFRs from \citet{chang15}, again split by galaxy type\footnote{We
  find that restricting the comparison to galaxies which
  \citet{chang15} flag as having good determinations has negligible
  effect on the results, so we use all galaxies regardless of that
  flag.}. Note that we normalize both SSFRs using our stellar
mass. Galaxies classified as star-forming on the BPT diagram (left
panel) compare well, but show a tail of anomalously low values of
Chang et al.\ SSFRs, which causes the best fit line to deviate from
unity. However, if the tail is excluded, the non-linearity disappears,
and a small bulk offset of 0.09 dex remains. We find that the offset
is entirely due to $\sim\!1/3$ galaxies that are not detected at 22
$\mic$ but only at 12 $\mic$, for which Chang et al.\ (S)SFRs are on
average 0.17 dex lower than our SED fitting (S)SFRs. For galaxies with
22 $\mic$ detection, the scatter with respect to SED SSFRs is 0.17
dex, compared to 0.23 dex between SED SSFRs and SSFRs from \wise\ 22
$\mic$ alone (Figure \ref{fig:comp_wise}, left panel). This reduction
of scatter demonstrates that the 12 $\mic$ photometry (a wavelength at
which \wise\ is significantly more sensitive than to 22 $\mic$) helps
constrain the SFRs compared to when 22 $\mic$ is used alone. However,
when 12 $\mic$ is used to obtain the IR luminosity (or SFR) without
the 22 $\mic$ measurements, it leads to a systematic underestimate.

For AGN hosts (the middle panel), the tail of anomalously low SSFR
values is more extensive than for the star-forming galaxies. We
confirm that these galaxies have UV detections, and that the UV
emission drives SSFR estimate in our SED fitting to higher values than
when the UV is omitted from the SED fitting, as in the case of
\citet{chang15} SED fitting. UV is not expected to be contaminated by
non-stellar emission in these, Type 2, AGN. Furthermore, the UV
emission is extended, suggestive of SF. Optical colors alone are not
sensitive to such low levels of SF \citep{kauffmann07}, and are
equally red for truly quiescent galaxies and those with intermediate
SSFRs \citep{fang12}. When the SED fitting is performed without UV
constraints, the SSFR defaults to very low values of the majority of
optically red models

For AGN hosts with high SSFRs, the excess seen in other IR-based SSFRs
is now smaller. This is true whether the galaxy was detected at both
12 and 22 $\mic$, or just at 12 $\mic$, suggesting that 12 $\mic$ flux
is much less affected by AGN-heated hot dust than the 22 $\mic$, so
that when included in the SFR estimate it mitigates the excess. That
22 $\mic$ suffers more AGN contamination is corroborated by the fact
that out of 12 $\mic$-detected AGN, 80\% are also detected at 22
$\mic$, whereas this fraction was 60\% for star-forming
galaxies. Previously, \citet{donoso12} have found that the 12 $\mic$
luminosity is not affected by AGN contribution in all but the handful
of most luminous sources (but they did not perform analogous
assessment of 22 $\mic$ emission).

For galaxies with weak or no \ha\ (right panel), \citet{chang15}
values agree well for some of the galaxies with high SSFRs, with a
small offset that we attribute to the preponderance (90\%) of 12
$\mic$-only detections in this category, which we have shown to have
somewhat underestimated SFRs in Chang et al. As in the case of
comparison with MPA/JHU catalog SFRs, the E+As, or more generally, the
post-starburst galaxies, again form a cloud of points offset from the
1:1 relation. As discussed in Section \ref{ssec:comp_b04}, the offset
in (S)SFRs is attributable to the differences in the assumed dust
attenuation laws, to which the E+As appear to be particularly
sensitive. Comparison with entirely independent IR SFRs as well as
better quality of SED fits, suggest that the high (S)SFRs obtained
with the modified attenuation law in GSWLC are more
realistic. Finally, for galaxies in ``no \ha'' class with log
SSFR$<-11$ the estimates largely agree, but are quite uncertain.

The general conclusion is that \citet{chang15} SFRs are reliable for
galaxies with log SSFR$_{\rm C15}>-10.7$, regardless of the galaxy
type, and especially when a galaxy is detected at 22 $\mic$. When
SSFRs are low, \citet{chang15} correctly attribute the mid-IR emission
to old populations and not the ongoing SF. However, in doing so, the
weak signal from the actual SF, detectable in the UV, is in some cases
lost, leading to anomalously low values of SSFR.

\citet{chang15} provide an estimate of the dust luminosity, i.e., the
total IR luminosity. In Figure \ref{fig:comp_lir_c15}, we compare it
to the IR luminosity that we derive from \wise\ 22 $\mic$. There is a
good overall agreement and the scatter is smaller than in the
comparison involving \citet{ellison16} $L_{\rm IR}$, no doubt because
\wise\ and Chang et al. IR luminosities are somewhat correlated
through the use of the same 22 $\mic$ photometry. There is, however, a
small fraction (1.8\%) of galaxies for which there is a $\sim\! 1$ dex
discrepancy in $L_{\rm IR}$.  These galaxies produce a bump in the
distribution of specific luminosities (log $(L_{\rm IR}/M_*$)) when
using \citet{chang15} IR luminosities, but not with \wise\
values. Most of these galaxies do not have $L_{\rm IR}$ from
\citet{ellison16} catalog, but when they do, their agree with our IR
luminosity from \wise. Finally, the direct comparison of Chang et al.\
IR luminosities with IR luminosities from {\it Herschel} (plot not
shown) confirms that Chang et al. IR luminosities are underestimated
when they are based on 12 $\mic$ detection without the 22 $\mic$
detection.

\section{Summary}

The paper presents GSWLC, an extensive catalog of physical parameters
(SFRs, dust attenuations and stellar masses) of $\sim$700,000 SDSS
galaxies covered by \galex\ (90\% of SDSS). SFRs and stellar masses
were derived using state-of-the art SED fitting of UV and optical
fluxes. SFRs derived independently from 22 $\mic$ \wise\ photometry
are also included in the catalog. 

The construction of the catalog, the internal checks and the
comparisons with previously published catalogs have produced a number
of results, which we summarize here:

\begin{enumerate}
\item The principal source of bias in \galex\ UV photometry is from 
  blending of unresolved sources. We provide empirical recipe to
  reduce its effect.

\item Total IR luminosities obtained from \wise\ 22 $\mic$
  observations via \citet{ce01} luminosity-dependent templates agree
  remarkably well ($\Delta<0.01$ dex, 0.07 dex of scatter) with IR
  luminosities obtained from \wise\ 22 $\mic$ and {\it Herschel}-SPIRE
  sub-mm bands.

\item The comparison of SED SSFRs with SSFRs from the mid-IR and,
  separately, with \ha\ SSFRs, suggests that the majority of galaxies
  require dust attenuation curve that is significantly steeper than
  the \citet{calzetti00} curve, and is on average similar to the
  \citet{conroy10uv} curve. Allowing this steep attenuation curve to
  include a UV bump further improves the quality of UV/optical SED
  fits. 

\item Not accounting for the emission lines in the SED modeling of
  broadband UV-optical photometry produces significant biases in the
  derived (S)SFRs: up to 0.5 dex, on average, for high-SSFR galaxies
  (log SSFR$>-9.5$).  Stellar masses are not affected by emission
  lines.

\item SFRs and stellar masses are not very different ($<0.1$ dex) when
  assuming a smooth, delayed SF history as opposed to the
  two-component (old and new) exponentially declining SF history,
  adopted for GSWLC. Fixing the old component to have started in the
  early universe ($t_0=10$ Gyr), as done here, yields 0.1 dex higher
  masses for galaxies with active SF, possibly removing the
  ``outshining'' bias.

\item For actively star-forming (``main'') sequence galaxies (log
  SSFR$>-10.5$), there is a good general agreement between SED fitting
  and mid-IR (S)SFR from GSWLC and also with (S)SFRs from the
  literature.

\item Aperture-corrected (total) SFRs from the MPA/JHU catalog
  \citep{b04} have no systematics at high SFRs when compared with any
  other SFR, in contrast to systematic offsets reported in
  integral-field spectroscopy studies. GSWLC, by offering two
  independent SFRs can potentially be used in the forthcoming
  integral-field spectroscopy studies to elucidate the source of the
  discrepancy.

\item IR luminosities of galaxies that host AGN (as identified from
  the BPT diagram) and have relatively high SSFRs, appear to have an
  excess IR emission, presumably due to AGN dust heating affecting the
  22 $\mic$ flux. If interpreted as SFR, this excess IR luminosity
  leads to SSFR overestimates of 0.2--0.6 dex, with greater effect at
  higher SSFR.

\item For galaxies that lie below the star-forming main sequence
  ($12<$log SSFR$_{\rm SED}<-11$), which includes quenching, nearly
  quiescent, or rejuvenated galaxies, SSFRs are low and their
  determination is challenging by any method. Using simple (fixed
  factor) conversions of IR luminosity to SFR (intended for use with
  actively star-forming galaxies) produces greatly exaggerated (S)SFRs
  (up to 2 dex). (S)SFRs from the UV/optical SED fitting tend to
  retain sensitivity in this regime. Below log SSFR$=-11.7$ (for
  GSWLC-M; $=-11.5$ for GSWLC-A, $-12.0$ for GSWLC-D), even the
  UV/optical SSFRs should be considered as upper limits, as these are,
  for all practical purposes, truly quiescent galaxies.
\end{enumerate}

SDSS is the workhorse dataset for many galaxy evolution studies at low
redshift, but without surveys in other wavelengths it has been limited
in terms of characterizing the SF.  We hope that GSWLC, by combining
SDSS with \galex\ and \wise, will fill this gap and serve as a
resource for many new discoveries.

\acknowledgments The construction of GSWLC was funded through NASA
ADAP award NNX12AE06G. We thank Sara Ellison for valuable comments. SS
thanks the members of \galex\ science team for their contributions
over the years, especially Tim Heckman, Chris Martin, R. Michael Rich
and Mark Seibert. Funding for SDSS-III has been provided by the Alfred
P. Sloan Foundation, the Participating Institutions, the National
Science Foundation, and the U.S. Department of Energy Office of
Science. The SDSS-III web site is http://www.sdss3.org/. Based on
observations made with the NASA Galaxy Evolution Explorer. GALEX is
operated for NASA by the California Institute of Technology under NASA
contract NAS5-98034. This publication makes use of data products from
the Wide-field Infrared Survey Explorer, which is a joint project of
the University of California, Los Angeles, and the Jet Propulsion
Laboratory/California Institute of Technology, funded by the National
Aeronautics and Space Administration.


\end{document}